\documentclass[conference]{IEEEtran}
\usepackage{amsmath}
\usepackage[ruled,vlined]{algorithm2e}

\usepackage{cite,cases,color,array,url}
\usepackage{algorithmic}
\usepackage{graphicx}
\usepackage{mdwmath}
\usepackage{mdwtab}
\usepackage{eqparbox}
\usepackage{amsopn}
\usepackage{multirow}
\usepackage{graphics}
\usepackage{subfigure}
\usepackage{url}
\usepackage{paralist}
\usepackage{xspace}
\usepackage{amsmath}
\usepackage{epsfig}
\usepackage{latexsym}
\usepackage{amsfonts}
\usepackage{amssymb}
\usepackage{comment}
\usepackage{mathrsfs}
\usepackage{cases}

\graphicspath{{./figures/}}

\def\ie{\textit{i.e.}\xspace}

\def\eg{\textit{e.g.}\xspace}

\newcommand{\eqqref}[1]{Eq.~(\ref{#1})}

\def\ourprotocol{{\textbf{C$^2$IL}}\xspace}

\DeclareMathOperator{\argmax}{argmax}

\begin{document}

\title{Communicating Is Crowdsourcing: Wi-Fi Indoor Localization with
  CSI-based Speed Estimation}

\author{\IEEEauthorblockN{Zhiping Jiang\IEEEauthorrefmark{1},
Jizhong Zhao\IEEEauthorrefmark{1}, Xiang-Yang Li\IEEEauthorrefmark{2},
Wei Xi\IEEEauthorrefmark{1},  Kun Zhao\IEEEauthorrefmark{1},
Shaojie Tang\IEEEauthorrefmark{3},
Jinsong Han\IEEEauthorrefmark{1}
}
\IEEEauthorblockA{\IEEEauthorrefmark{1}School of Electronic and
  Information Engineering, Xi'an Jiaotong University, Xi'An, China}
\IEEEauthorblockA{\IEEEauthorrefmark{2}Department of Computer Science,
  Illinois Institute of Technology, Chicago, IL}
  \IEEEauthorblockA{\IEEEauthorrefmark{3}Department of Computer and Information Science,
  Temple University, Philadelphia, PA}
Email: \{\emph{jiangzp.cs, weixi.cs, tangshaojie }\}@gmail.com,
\emph{xli}@cs.iit.edu, \{\emph{zjz, zhaokun, hanjinsong}\}@mail.xjtu.edu.cn}

\maketitle

\begin{abstract}
Numerous indoor localization techniques have been proposed recently to
 meet the intensive demand for location based service, and Wi-Fi
 fingerprint-based approaches are the most popular and inexpensive
 solutions.
Among them, one of the main trends is to incorporate the
 built-in sensors of smartphone and to exploit crowdsourcing
 potentials.
However the noisy built-in sensors and multi-tasking limitation of
 underline OS often hinder the effectiveness of these schemes.

In this work, we propose a \textit{passive \underline{c}rowdsourcing}
\underline{C}SI-based \underline{i}ndoor \underline{l}ocalization
scheme, \ourprotocol.
Our scheme \ourprotocol \emph{only} requires the locating-device (\eg,
 a phone) to have a 802.11n wireless connection,
 and it does not  rely on inertial sensors only existing in some
 smartphones.
\ourprotocol is built upon our innovative method to accurately
 estimate the moving distance purely based on 802.11n Channel State
 Information (CSI).
Our extensive evaluations show that the moving distance estimation
 error of our scheme is  within $3\%$ of the actual moving distance
 regardless of varying speeds and environment.
Relying on the accurate moving distance estimation as constraints,
 we are able to construct a more accurate mapping between RSS
 fingerprints and location.
To address the challenges of collecting fingerprints,
 a crowdsourcing-based scheme  is designed to \emph{gradually} establish the
 mapping and populate the fingerprints.
In \ourprotocol, we design  a trajectory clustering-based localization
 algorithm  to provide precise real-time indoor localization and
 tracking.
We developed and deployed a practical working system of \ourprotocol
 in a large office environment.
Extensive evaluation results indicate that our scheme \ourprotocol
  provides accurate localization with error
  $2m$ at $80\%$ at very complex  indoor environment with minimal
  overhead. 

\end{abstract}

\section{Introduction}
\label{sec:introduction}

With the prosperity of mobile devices, especially smartphones, 
 location based services (LBS), which use the geographic
 position to provide targeted services, have become
 pervasive to provide added value of existing services.
A critical challenge of LBS is to find the accurate location of
 mobile devices.
GPS  has successfully dominated the outdoor localization.
Unfortunately in indoor environment, the most facile wireless received
 signal strength (RSS) is neither accurate nor consistent due to the
 highly dynamic and complex environment.
As the flourishing of smartphones and crowdsourcing computation models,
 numerous indoor localization techniques have been proposed to collect
 fractional environment features and collaboratively provide
 precise indoor localization.

The relatively good accuracy and simplicity of fingerprint based
 localization schemes has attracted massive of effort in the
 community.
Wi-Fi fingerprint-based schemes can provide
 meter-level  indoor localization accuracy at the expense of explicit 
 site-survey.
Its high deployment cost and low adaptiveness to environment
 change  hinders the practical effectiveness.
Recently several novel techniques, \eg,
 \cite{spinloc,FILA,Zee,centaur,pushingLimit,lifs},
 have been proposed to raise the usability and accuracy.
Among these approaches, a hot research trend is to incorporate
 crowdsourcing model and built-in sensors in today's
 smartphone.
LiFS~\cite{lifs} reduces the site-survey by using the
 moving distance,  estimated from counting the number of steps by
 accelerometer, as constraints for matching between the map and
 trace-graph, achieving average accuracy of $5.8m$ 
Zee~\cite{Zee}  achieves a mean accuracy $3m$ by estimating the moving
 direction and   moving distance by similarly leveraging the sensors
 in smartphone.
Centuar~\cite{centaur} and PAL~\cite{pushingLimit}
 both calibrate the Wi-Fi fingerprints database using acoustic ranging
 and they achieve $1$ to $3m$ accuracy.
All these  approaches significantly improved the practicability of
 Wi-Fi based indoor localization, however, we believe there are plenty
 of room for improving the localization accuracy while
 reducing or even eliminating the dependence of site-survey and noisy
 inertial  sensors. 

In this work, we design and develop an indoor
 localization scheme \ourprotocol with even lower cost and hindrance.
Our scheme  exploits the  channel state
information (CSI) in 802.11n, for better
 distance estimation which in turn results in a 
 better fingerprints matching  and localization.
Additionally, \ourprotocol, as in most recent techniques,
  collects the WiFi RSS fingerprints  during
 communication between the user and WiFi APs, and populates the
 fingerprints database by crowdsourcing.
 \ourprotocol performs extremely well in a complex
 environment with rich multipath effect, while several recently
 developed schemes suffer from lower localization accuracy in such a
 complex environment.

\begin{figure}[tb]
\begin{center}
\begin{tabular}{cc}
\hspace{-0.1in}
\includegraphics[scale=0.49]{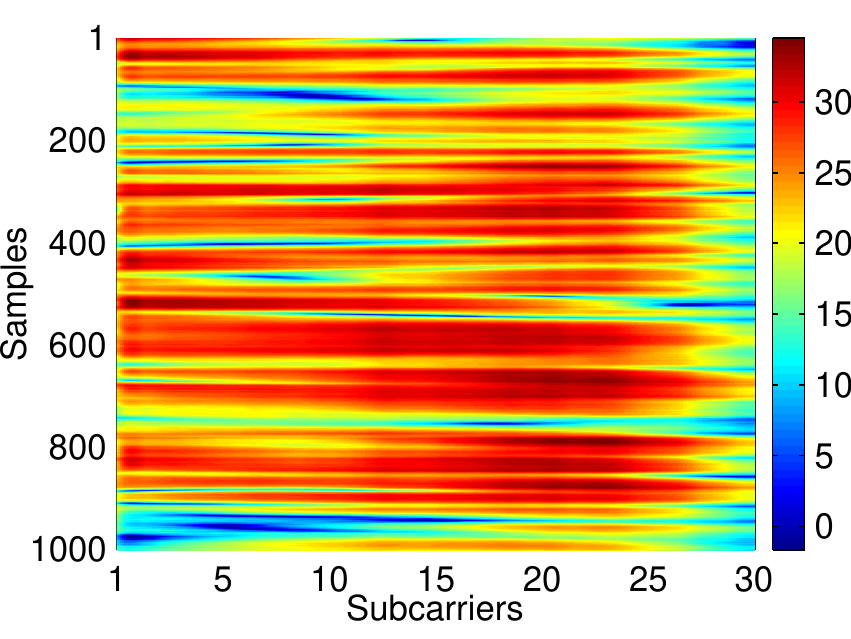} & \hspace{-0.15in}
\includegraphics[scale=0.49]{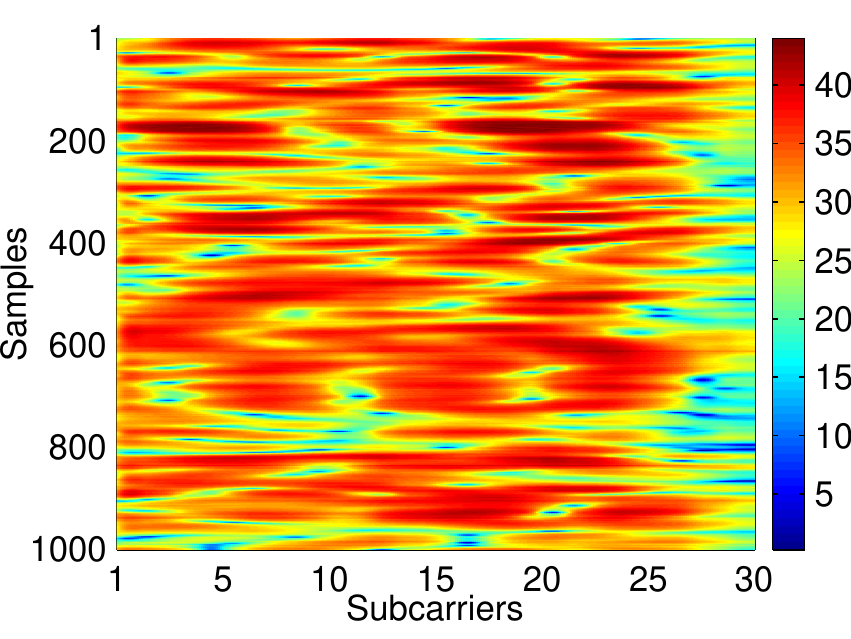}  \\
(a) Ripples in SISO & (b) Ripples in MIMO
\end{tabular}
\vspace{-0.15in}
\caption{(a) Ripples across nearly all subcarriers in SISO  (b) The
  ripples in $3\times 3$ MIMO configuration with MCS=16.}
\label{fig:csisample}
\vspace{-0.2in}
\end{center}
\end{figure}

The accurate moving distance estimation is based on the readily
 available  CSI in IEEE 802.11n systems that use OFDM MIMO technology.
Since CSI focuses more on  small-scale fading, it has extraordinary
 advantages on capturing the mobile channel characteristics.
From our preliminary testing on various scenarios when users move with
 constant speeds in a complex environment with rich multipath features,
 we find ripples-like Rayleigh deep
 fadings~\cite{rappaport1996wireless} with some periodicity across all
 subcarriers regardless of antenna configurations as shown in
 Fig.~\ref{fig:csisample} (a) and (b). A natural question is
``\textit{Are these ripples correlated with distance or speed}?''.

The answer is beyond a simple \emph{yes}.
In this work, we first exhibit a simple yet precise relationship
 between moving speed $v$ and the frequency of ripples-like fadings
 $f_r$.
Based on this important discovery, we for the first time propose a
 simple method to precisely estimate the moving speed (and moving
 distance) of a wireless device purely based on wireless traffic
 between the device and the WiFi APs.
Since in Wi-Fi environment both AP and clients can measure the CSI and
 RSS, the client's moving speed and RSS can be
 measured remotely at AP-end \emph{passively} and \emph{silently}.
In such a case, all the connected client devices have already become
 participants of the crowdsourcing system without any effort.
When there is communication, they are formally contributing to the
 crowdsourcing, so \emph{communication is crowdsourcing} in \ourprotocol.

After computing a precise speed and moving distance estimation,
 our localization scheme \ourprotocol  then applies the graph matching
 (GM) for several core tasks: fingerprints extraction, automatic
 mapping between fingerprints and floor plan, and
 user localization/tracking.
It has been widely known that the RSS value is affected by many
 factors, \eg, the RSS values collected at the same location using same
 devices with same WiFi APs could fluctuate to a few db depending on
 how users hold and block the signal \cite{spinloc}.
Such fluctuation will significantly impact the fingerprint matching
 quality and thus impact the localization/tracking accuracy.
We first identify and successfully address the
 directional shadowing problem for
 conventional threshold-based fingerprints extraction, and we propose
 a trajectory matching-based solution to eliminate the shadowing problem.
For  of fingerprints mapping task, \ourprotocol supports
 unsupervised large scale complex indoor floor plan mapping.
For  localization/tracking task, a combination of
 trajectories matching and particle filter is proposed to achieve
 precise indoor localization/tracking.
In summary, the contributions of \ourprotocol are as follows.

$\bullet$ 
Compared with the previous approaches with good accuracy, our scheme
\ourprotocol
 does not require the localization devices to be smartphones with
 various inertial sensors (which are required in \cite{Zee,lifs}).
As it only requires  802.11n, \ourprotocol has the lowest barrier on
 devices, thus, we expect it to have better contribution from users,
 which is of vital importance for  practical and
 continuously-functioning crowdsourcing.

$\bullet$  \ourprotocol is the first
practical  localization system that really benefits from the
multipath effect, instead of suffering from this notoriously challenging
effect. Theoretical analysis showed that it is exactly the
multipath effect that enables the accurate distance estimation by
CSI. Our extensive experimental evaluation  indicated that in typical indoor
scenarios, the distance estimation error is often within 3\% 
 regardless of moving speeds, which is much more
 accurate than Dead-Reckoning or pedometer based approaches.

$\bullet$ The adoption of graph matching and other techniques in our
core design guarantees the accuracy and scalability of RSS map
building and localization in very large and complex environment. We
design and develop a prototype of \ourprotocol in a large office
environment of about 2000$m^2$ with complex structure.
In our extensive tests, the localization error without any historical
 data is within $5m$; while  during tracking, thanks to the precise
 moving distance estimation by  CSI, the tracking error could be within $1m$.

The rest of the paper is organized as follows.  
We review  related techniques in Section~\ref{sec:related},
 present  \ourprotocol system overview in Section~\ref{sec:overview}
 and our innovative distance estimation in
 Section~\ref{sec:estimating}. 
Fingerprint extraction and complete
  crowdsourcing based localization scheme are introduced in
Section~\ref{sec:fingerprintExtraction} and
Section~\ref{sec:crowdsourcing} respectively.
We present the fine-grained indoor tracking  in
Section~\ref{sec:tracking}, report our extensive performance
evaluation of \ourprotocol in Section~\ref{sec:evaluation}.
 We conclude the paper in Section~\ref{sec:conclusion}.

\section{Related Work}
\label{sec:related}

\subsection{Indoor Localization Schemes}

The localization problem are applied in two main scenarios: outdoor
and indoor.
The most popular outdoor localization method is GPS~\cite{liu2012energy}.
Other existing techniques of both outdoor and indoor localization
 mainly fall into two categories: Fingerprint-based and
 Modeling-based.
Fingerprints are utilized in many literatures to assist positioning,
 and the most widely used one is WiFi signal.
In indoor environment, fingerprint based methods (\eg, Radar
\cite{bahl2000radar}, Horus~\cite{youssef2008horus},,
SurroundSense~\cite{azizyan2009surroundsense},
PinLoc~\cite{sen2012you}, )
 first collect  fingerprint of WiFi signal (or cellular, or FM, or
 other sensors such as light)
 in advance at known locations inside a building, and then identify
 the user's location by matching the fingerprint of this user with the
 fingerprint stored in database.
Dead-Reckoning is another stream of techniques (\eg,
\cite{constandache2010towards,guha2010autowitness})
 proposed in the literature for localization.

The most used fingerprint is the RSS value.
LiFS \cite{lifs} proposed a crowdsourcing based indoor localization, which
 exploits the possibility of automatically establishing the mapping
 between fingerprint set $F$ and position set $P$.
Acoustic ranging (AR) assisted Wi-Fi positioning was recently
developed to provide distance estimation between two users (\eg,
 \cite{centaur,pushingLimit}).
These schemes leverage the accurate AR and
 are able to provide high localization accuracy using the mapping of
 fingerprints with some additional distance constraint.

CSI has potential for accurate indoor localization
 since the CSI tool\cite{csitool} has been released to public on
 off-the-shelf  hardware.
CSI is  not a simple extension of RSS on physical
 subcarriers but it reveals totally different information on frequency
 selective fading process.
SpinLoc \cite{spinloc} system proposed a
 rotation based indoor localization system that leveraged the human
 bodies' strong fading to Line-of-Sight (LoS) components.
PinLoc~\cite{pinloc} proposed a CSI fingerprint-based localization
 system which can achieve meter-level precise indoor point
localization.
FILA~\cite{FILA} proposed a precise indoor ranging system by
 eliminating  the non-LoS components in CSI information.
Compared with these schemes, our scheme \ourprotocol provides an accurate
 moving distance estimation of a single user in a complex indoor
 environment.

\subsection{Estimate Moving Distance by CFR}

There is a long history of estimating moving velocity of a mobile
 station according to wireless signal\cite{azemi2004mobile,mohanty2005vepsd}.
Most  of them focused on fast moving stations, \ie, a mobile station in cars
 or trains.
However, the algorithm for estimating the maximum Doppler
 frequency $f_d$, on which most of these methods based,
 is not suitable for estimating human walking speed.
The maximum Doppler frequency in
 2.4G or 5G Wi-Fi environment could be almost totally ignored, \ie, for
 a moving station with a velocity $1.5m/s$, $f_d$ is merely 12Hz/25Hz
 compared to the carrier frequency 2.4Ghz or 5.2Ghz.

To the best of our knowledge, \cite{speedestimation1} is the only
 previous work that implemented an indoor speed estimation system, which
 is based on DVB-T signal working at 746Mhz. This work used the
relationship $v=\xi\frac{\lambda}{T_c}$ to estimate velocity, where
$\xi=0.423$ a pre-defined constant, $\lambda$ the wavelength, and
$T_c$ is the channel coherent time. However, for Wi-Fi signal with
small covering range, which causes non-uniformly Rayleigh fading, a
constant $\xi$ is not appropriated, and it is extremely difficult and
challenging to precisely estimate the $\xi$ in a dynamic indoor
environment.

\subsection{Graph Matching via Relaxation}

Graph matching (GM) is a widely used technique to find the best
 correspondence between two graphs, documents, or images. In
 \ourprotocol, we use GM to extract RSS fingerprints and  automatically
 establish the mapping between RSS fingerprints and floor plan.

GM is essentially an integer quadratic programming (IQP) problem,and it
 is NP-hard~\cite{cho2010reweighted}.
Given two graphs $G^P=(\mathcal{V}^P,\mathcal{\varepsilon}^P)$ and
$G^Q=(\mathcal{V}^Q,\mathcal{\varepsilon}^Q)$, the goal of GM is to
find the best correspondence between  two graphs. Let $C^{P\times
  Q}$ represent the possible matching candidates set, the affinities
 between all candidates are recorded in an adjacency matrix
$\mathcal{M}^{C^{P\times Q}}$  based on applications.
Let $\mathcal{X} \in \{0,1\}^{n^\mathcal{M} {C^{P\times Q}}}$ be a
column-wise binary vector which indicates the selected
correspondences, the graph matching problem can be expressed as
finding the best indicator vector $\mathcal{X}^*$ that maximizes a
score function $S(\mathcal{X}) =\mathcal{X}^T\mathcal{M}^{C^{P\times
    Q}}\mathcal{X}$, \ie,
 $\mathcal{X}^* = \argmax S(\mathcal{X})$.

Various relaxation based methods have been proposed. Most of them
relax the integer constraint of $\mathcal{X}$.
After obtaining the
 optimal $\mathcal{X}^*$ in real number domain with different insights
 of $\mathcal{M}$, the discretization of $\mathcal{X}^*$ will make the
 best approximated solution to the underlined IQP problem.
In \ourprotocol, two major graph matching algorithms, spectral
 matching (SM)~\cite{leordeanu2005spectral} and RRWM
 algorithm~\cite{cho2010reweighted} are used in different stages.
SM is used in extracting RSS fingerprints graph; while RRWM is used in
mapping between RSS fingerprints and floor plan.

%
%

\section{Architecture Overview}
\label{sec:overview}
Since CSI and RSS can be estimated by both AP and clients,
\ourprotocol  can be deployed in either AP-end or client-end. Whenever
it is deployed at either end, the core of the system remains. The
positions of APs are not required to be known in either case.
%
%
%

\begin{figure}
\begin{center}
\includegraphics[scale=0.5]{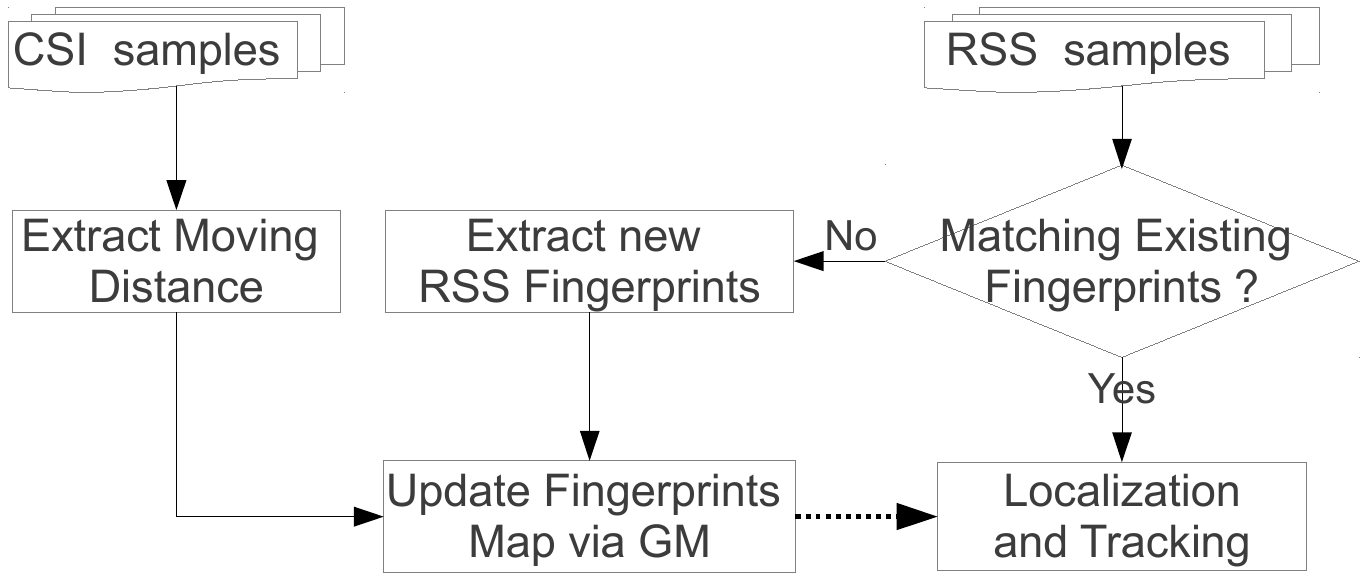}
\end{center}
\vspace{-0.1in}
\caption{\ourprotocol  System Architecture.}
\vspace{-0.1in}
\label{fig:overview1}
\end{figure}

Fig.~\ref{fig:overview1} demonstrates our system architecture.
When a user enters the building, we assume that the device held by the
 user will build wireless connections with the APs inside the building.
The APs will record the RSS values and the CSI values of the received
 signal from the client device and send the data to a localization
 server.
Based on the sequence of the CSI values taken during a time-window,
 the server  will  then quickly estimate the moving distance of the
 client in this  time-window.
The RSS values will be used to build a matching with the fingerprint
 RSS values  stored in the fingerprint database, which itself is
 populated using crowdsourcing techniques.
The estimated moving distance, together with the estimated geodesic distance of
different fingerprint locations  in the map, will be used to further improve the
quality of the matching, and thus, the accuracy of the localization.


%
%

\section{Estimating Moving Speed by CSI}
\label{sec:estimating}
In this section, we mainly focus on the techniques of estimating
moving speed and thus moving distance by CSI.
Theoretical basis is presented first, and then
 we present our algorithm implementation.

\subsection{The Electromagnetic Standing Wave Field}

Wireless radio propagation in compact environment could be modelled as a
superposition of large-scale path-loss, middle-scale shadowing, and small-scale
multipath fading~\cite{rappaport1996wireless}. For the multipath fading, it is
usually fitted to a statistical model called Rayleigh or Rician (Rayleigh fading
plus strong LoS components) distribution. The ripples-like deep fading shown in
Fig.\ref{fig:csisample} (b) and (c) are typical Rayleigh fading pattern.

Previous speed estimation methods, are based on some statistical properties of
Rayleigh distribution, \eg level crossing rate (LCR) or coherent time $T_c$.
Although it has been experimentally validated that the distance between two
adjacent ripples (deep fadings) is about $\lambda/2$ ($\lambda$ is the carrier
wavelength) even in large-scale multipath environment (like Manhattan city)
\cite{sklar1997rayleigh}, no previous works explicitly exploits such $\lambda/2$
fluctuation, since such fluctuation is encapsulated and blurred in a too general
model.

However, some detailed studies of radio
propagation\cite{braun1991physical,zonoozi1996shadow,chen1997sbr} have indicated
that, in a complex multipath environment the constructive or destructive
interferences of the large sum of reflected and scattered waves will generate a
\textit{standing waves field}, and the environment becomes a weak
\textit{Electromagnetic Cavity Resonator} (ECR) \cite{hill2009electromagnetic}
which hold standing waves in a very short time. According to basic physics of
wave propagation, the distance between two adjacent \textit{antinodes} (position
with maximum amplitude), \textit{towards any direction}, is $\lambda/2$, thereby
the experimentally observed $\lambda/2$ fluctuation.
Therefore, when a antenna traverse the indoor space with a speed $v$, a
periodically ripples-like pattern with a frequency $f_o = 2\frac{v}{\lambda}$
appears. Such simple relationship inspires us that the moving speed $v$ could be
precisely estimated purely from the CSI, if we could precisely estimate $f_o$.

\subsection{Theoretical Basis}
Wireless signal propagation in indoor environment can be well modelled
as Rician fading channel. In such a model, the CFR at the $i$-th
subcarrier is described as \cite{ricianKestimate}
\begin{equation}
H_{rice}(i)=\sqrt{\dfrac{1}{K+1}}H_m(i)+\sqrt{\dfrac{K}{K+1}}H_{LoS}(i)
\end{equation}
where the $H_{LoS}(i)$ represents the deterministic Line-of-Sight (LoS)
component, $H_m(i)$ represents the random multipath component. The
Rician factor $K$ determines the power ratio between these two
components.

To simplify the system model, we will first focus on a simplified
model, Rayleigh fading channel, which is a specialized form of Rician
fading channel when $K=0$. Generalized solution under Rician channel
will be discussed later.

\subsubsection{In Rayleigh Fading Channel}

In a multipath environment, there are lots of objects in the
environment that scatters the wireless signal. The received fading
envelop at an antenna will be the superposition of a large number of
these reflected and scattered waves. Since different position has
different constructive or destructive interference pattern among these
waves, the received signals amplitude at different locations become a
random variable. When a mobile antenna passes through the environment,
ripples-like deep fading appears in the instantaneous CFR as shown in
Figure \ref{fig:csisample} (b) and (c).

Assuming Wide Sense Stationary Uncorrelated Scattering (WSSUS)
 environment and the uniformly distributed angle of arrival (AoA) of
 multipath components, Clark \cite{Clark} has derived the
 auto-correlation of Rayleigh faded CIR $h(t)$ with motion at a scalar
 velocity $v$ \textbf{towards any direction} is a \textit{zeroth-order Bessel
function
  of the first kind} that
\begin{equation}
E\{h(t)*h(t+t_0)\}=J_0(2\pi f_dt_0)
\label{eq:bessel1}
\end{equation}
where $J_0(2\pi f_dt_0)$ is the bessel function with delay $t_0$ when the
maximum Doppler shift is
 $f_d$.
A general definition of the Bessel function, for integer values of $n$
 could be represented in an integral form:
\begin{equation}
J_n(x) = \dfrac{1}{\pi}\int_{0}^{\pi}\cos(nt_0 -x\sin t_0)dt_0
\end{equation}
For $J_0(x)$, the function can be further simplified to
\begin{equation}
J_0(x)\approx\sqrt{\dfrac{2}{\pi x}}\cos(x-\dfrac{\pi}{4})
\label{eq:bessel3}
\end{equation}
There is a large positive value $\epsilon$ , when $x>\epsilon$,
\eqqref{eq:bessel3} can be approximated by a periodical function
that
\begin{equation}
J_0(x)= J_0(x+2\pi), x>\epsilon,\epsilon\gg 0
\label{eq:bessel4}
\end{equation}
Substituting \eqqref{eq:bessel1} into Equation \eqqref{eq:bessel4}, we get:
\begin{equation}
J_0(2\pi f_dt_0)= J_0(2\pi f_d(t_0+\dfrac{1}{f_d}))
\label{eq:bessel5}
\end{equation}
Substituting the maxium Doppler frequency $f_d = \dfrac{v}{\lambda}$
into  \eqqref{eq:bessel5}, we obtain:
\begin{equation}
J_0(2\pi\dfrac{v}{\lambda}t_0)=J_0(2\pi\dfrac{v}{\lambda}(t_0+t_\lambda))
\label{eq:bessel6}
\end{equation}
where $t_\lambda = \dfrac{\lambda}{v}$.

Apparently, $t_\lambda \times v=\lambda$, and  \eqqref{eq:bessel6}
tells a very elegant result that, \textbf{iff} $\int_{t_1}^{t_2}vdt =
\lambda/2$,
\begin{equation}
J_0(2\pi f_dt_1)= J_0(2\pi f_dt_2)
\label{eq:bessel7}
\end{equation}

The periodicity of the auto-correlation function of $h(t)$ will
undoubtedly reflect on the original CIR $h(t)$, and CFR
$H(t)$. Thereby we could say, in a typical Rayleigh fading dominant
isotropically scattering environment, the distance $d_f$
between two adjacent deep fading(s) is a half-wavelength $\lambda/2$.
For example, In a 2.4G channel the wavelength $\lambda_{2.4G}\approx
12.5cm$, and in 5G channel the wavelength $\lambda_{5G}\approx 5.4cm$.

Thus, for the purpose of estimating moving distance,
 if there is a method to accurately count the number of deep fading(s)
 $N_f$ during a time interval $\Delta t$, we are able to precisely
 estimate the moving distance $S$ by
\begin{equation}
N_f\times \lambda=S
\label{eq:distance1}
\end{equation}

Noticing that the error caused by Doppler effect does exist in the
estimated distance $S$.
However since Doppler frequency caused by human
 walking $fd$ is in tenth level, which is too small compared to the
 channel frequency.
Thus, the error caused by Doppler effect is ignored.

\subsubsection{In Rician Fading Channel}
distinction between generalized Rician fading and Rayleigh fading.
The ripple-like deep fading is caused by strong interference
 (constructive or destructive) in multipath environment. When In strong rician
fading environment,
   high $K$ value will
 weaken the multipath effect and make it difficult to recognize and
 count the ripples.

\begin{figure}[t]
\begin{center}
\includegraphics[scale=0.9]{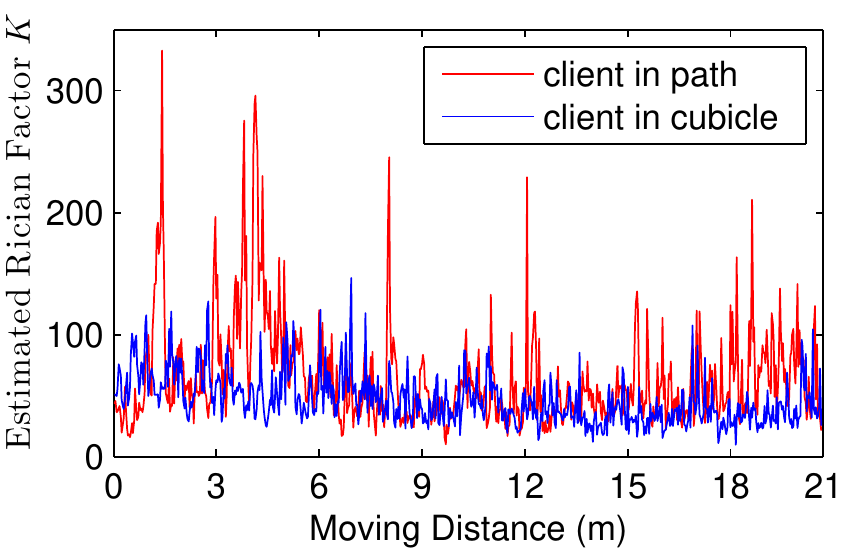}
\end{center}
\caption{Estimated Rician Factor $K$ during the test, where the mobile
  AP is moving towards the end. Higher $K$ denotes stronger LoS
  components reception. The client in a path has always-LoS connection,
  while there is no LoS component for the client in the metal-framed
  cubicle.}
\label{fig:ricianfactor}
\end{figure}

However,according to our experiments in typical indoor environment, we found
that even in a
 always-LoS path, high value $K$ rarely happens.
Figure \ref{fig:ricianfactor} plot the  $K$ values estimated from two
clients during the test using a moment-based estimator according to
\cite{ricianKestimate}.
The client in the path has always-LoS connection,
 while the LoS component is cut-off for the client placed in
 a metal-framed cubicle.

Out of our expectation, comparing to the client in cubicle, the $K$ value
 measured by the client in the path doesn't benefit much from the
 always-LoS connection.
Very large $K$ values only happen occasionally at the
 beginning, while it increases very slowly even when the AP is quite
 close to the end.
The occasional large values $K$ at prophase is simply caused by the indoor
structure at these locations, where additional propagation path is
introduced. These additional paths dilute and absorb some of the
multipath component.
Since the large value $K$  rarely happens even in the condition with
LoS, the distance measured using the Rician fading could be approximated
according to the moving velocity before and after the Rician fading.

\subsection{Speed Estimation Algorithm}

As previously described, the speed estimation problem now becomes a
specific frequency estimation problem. We design a reasonable and effective
processing flow.  It includes Data Preprocessing, Noise Cancellation, Fading Enhancement, and Frequency Estimation.
the oscillation frequency $f_o$.

\textit{Data Preprocessing:}
Every frame sent in 802.11n MCS rate at time $t$ has an CSI $H_t$. It
is a complex-number vector with a length of $N_{ss}\times L_{ss}$,
where $N_{ss}$ and $L_{ss}$ are the number of MIMO spatial streams and
number of measured subcarriers across the Wi-Fi bandwidth. Every
complex value $h^i_t \in H_t$ describes the instantaneous amplitude
$a^i_t$ and phase $\theta^i_t$ of the underlying $i$-th subcarrier.
In order to enable all 802.11n compatible devices to be ready for
speed estimation, we only use the first  spatial stream (first
$L_{ss}$ complex values in $H_t$) to estimate speed. Moreover, the
computation space is greatly reduced.
Since in multipath environment the phase $\theta^i$ is uniformly
distributed between $[0, 2\pi]$\cite{rappaport1996wireless}, which
provides no discriminative information.
Thus we drop the phase $\theta^i_t$
and only use amplitude $A=\vert H \vert$ to estimate speed.

The amplitude matrix $\mathbf{A}_{ori}=\{A_1,...,A_n\}^T$ is further
defined, where $A_i$ is the $i$th received column-wise
amplitude. Since the instantaneous reception rate of frames is
unstable due to the wireless traffic control, $A_{ori}$ is resampled
to a stable reception frequency $f_w$ with the even interval between
each slot, and let $\mathbf{A}_{re}$ denote the resampled amplitude
matrix.

\textit{Noise Cancellation:}
Convolution based noise cancellation is applied on $\mathbf{A}_{re}$
to filter out the high frequency noise, that
$\mathbf{A}_{nc}=\mathbf{A}_{re}\ast h_b(r)$,
\[h_b(r) =r\cdot\mathbf{1}^s\]
where $\mathbf{1}^s$ is a full-$\mathbf{1}$ square matrix of size
$s$. Currently in our system $s=6$. This step is of great importance
according to the real data evaluation, since the following fading
enhancement and frequency estimation is quite sensitive to noise.

\begin{figure}[t]
\begin{center}
\includegraphics[scale=0.5]{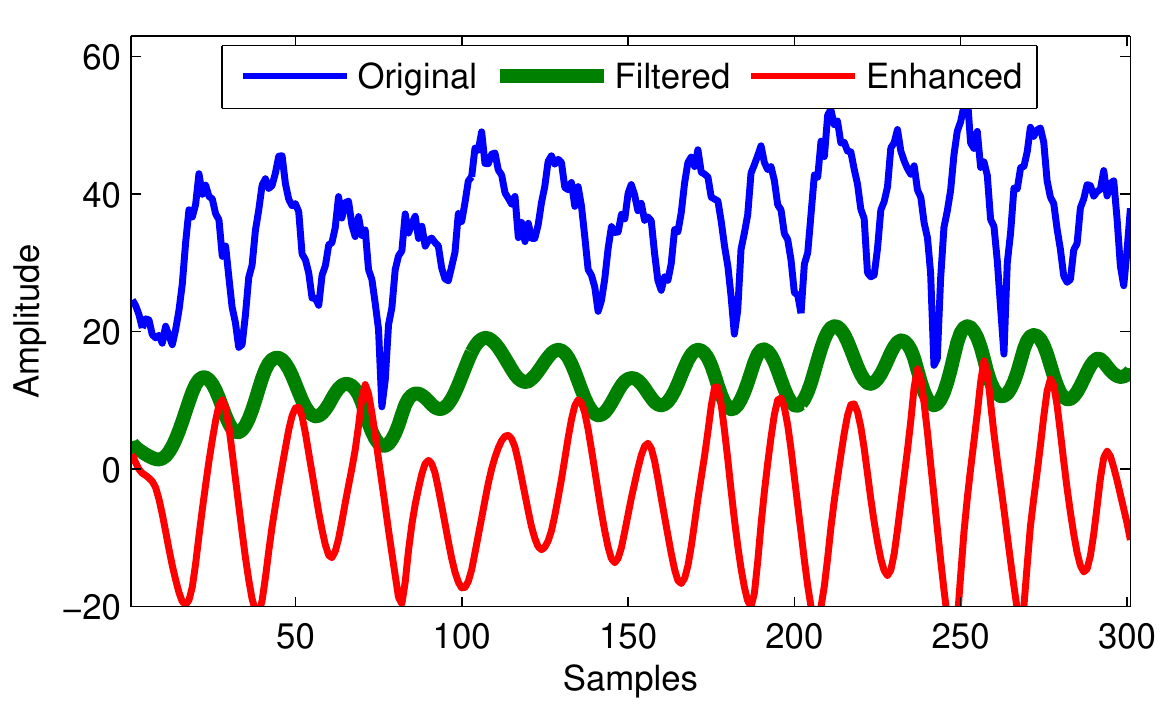}
\vspace{-0.15in}
\caption{The effect of each step of processing.}
\label{fig:power}
\vspace{-0.2in}
\end{center}
\end{figure}

\textit{Fading Enhancement:}
An intuitive idea of enhance the fading is first-order
derivation of $\mathbf{A}_{nc}$, however, first-order
derivation is quite sensitive to high frequency noise
rather than low frequency ripples. 
Another convolution is used to emphasize the fading that $\mathbf{A}_{en} =
\mathbf{A}_{nc} \ast h_{df}$, where $h_{df}$ is a Sobel-style
calculator that
\begin{equation}
h_{df} =
\begin{bmatrix}
2 & 5 & 2 \\
0 & 0 & 0\\
-2 & -5 & -2
\end{bmatrix}
\label{eq:sobel}
\end{equation}
Fig.~\ref{fig:power} shows the intermediate results After first 3 processing, and it is now suitable for frequency estimation.

\textit{Frequency Estimation:}
Due to the MIMO configuration or other interference, deep fading(s) in
all subcarriers are not guaranteed to appear simultaneously, as shown
in Fig.~\ref{fig:csisample}, therefore the final decision of $f_o$ are
based on the estimations of each underlying subcarriers.

Extracting $f_o^i$ for $i$-th subcarrier is equivalent to extracting
the expected frequency $E(f^i_c)$ in the spectral graph within a
frequency interval $f_{min}<f_c<f_{max}$. According to
Eq. \eqref{eq:speed}, $f_{min}$ and $f_{max}$ are set according to the
speed interval of human walking that
\begin{equation}
f_{min} = 2\cdot v_{min}/\lambda, \quad
f_{max} = 2\cdot v_{max}/\lambda
\end{equation}
where the minimum speed $v_{min}$ and the maximum speed $v_{max}$ in
our system are set to 0.8m/s and 1.6m/s.

Short-Time Fourier Transformation (STFT) with 50\% overlapping window
is applied to obtain the Power Spectral Density (PSD) of $i$-th
envelope of $A_{re}$. It reveals the spectral density of subcarrier
$i$ along with time. To reduce the jitter, the estimated $f^i_o$ is
set to the weighed expectation of frequencies between $f_{min}$ and
$f_{max}$
\begin{equation}
f^i_o = \frac{\sum_{f_{min}<s_j<f_{max}} s_j\cdot w_j}{\sum w_j},
\end{equation}
where $w_j$ denotes the power of frequency $f_j$.

\begin{figure}[tb]
\begin{center}
\includegraphics[scale=0.67]{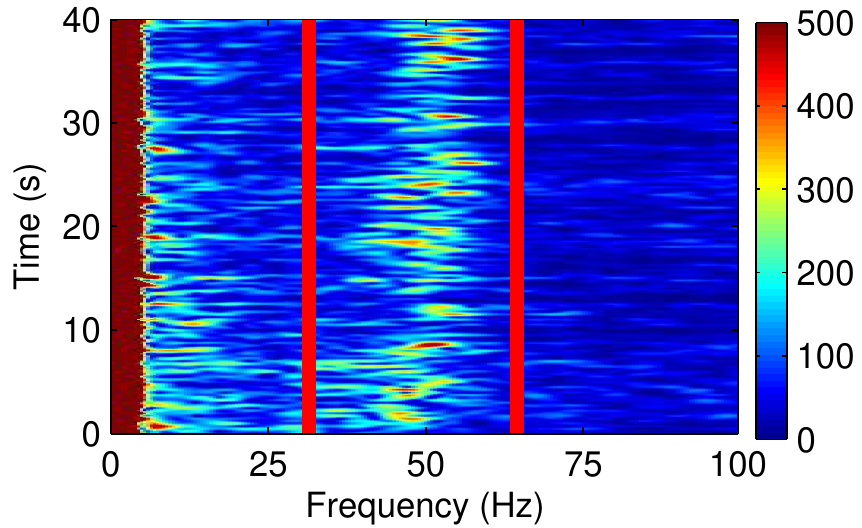}
\end{center}
\vspace{-0.25in}
\caption{STFT result for 15$th$ subcarrier at 5.8Ghz (Channel 161).}
\label{fig:stft}
\vspace{-0.1in}
\end{figure}

Figure \ref{fig:stft} shows the STFT result for $15$-th subcarrier.
We can see very strong power around
50Hz. As $50\times\frac{\lambda}{2}=1.29m/s$, and it is quite close to
 the real walking speed at about $1.3m/s$. Two red bars denote the
$f_{min}$ and $f_{max}$, and they are set to 32 and 64 according the
$v_{min}$ and $v_{max}$ settings.
The final estimation of $f_o$ is set to the median of all estimated $f_o^i$
\begin{equation}
f_o = median(f^1_o,f^2_o,...f^n_o), n=N_{ss}
\end{equation}
Then the moving speed   is estimated by
\begin{equation}
v = \frac{\lambda\cdot f_o}{2}
\label{eq:speed}
\end{equation}

It was worth to mention that although the Doppler effect does exist,
as discussed in Section \ref{sec:related}, the Doppler frequency is
 small comparing to the carrier frequency. Thus in current
system design, we did not consider the Doppler effect caused
by human walking.

\subsection{Start/Stop Detection}

\begin{figure}[t]
\begin{center}
\begin{tabular}{cc}
\includegraphics[scale=0.45]{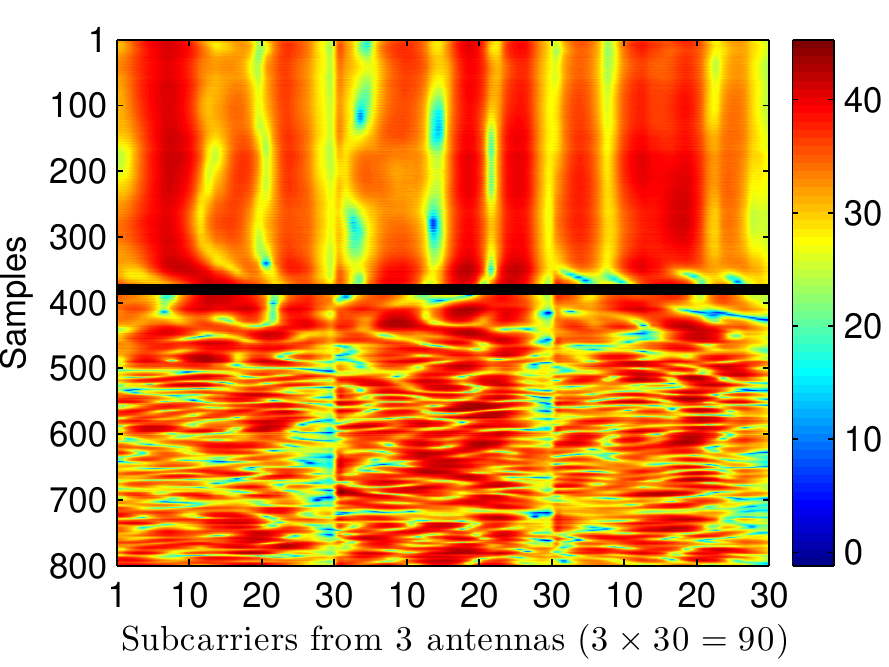} &
\includegraphics[scale=0.45]{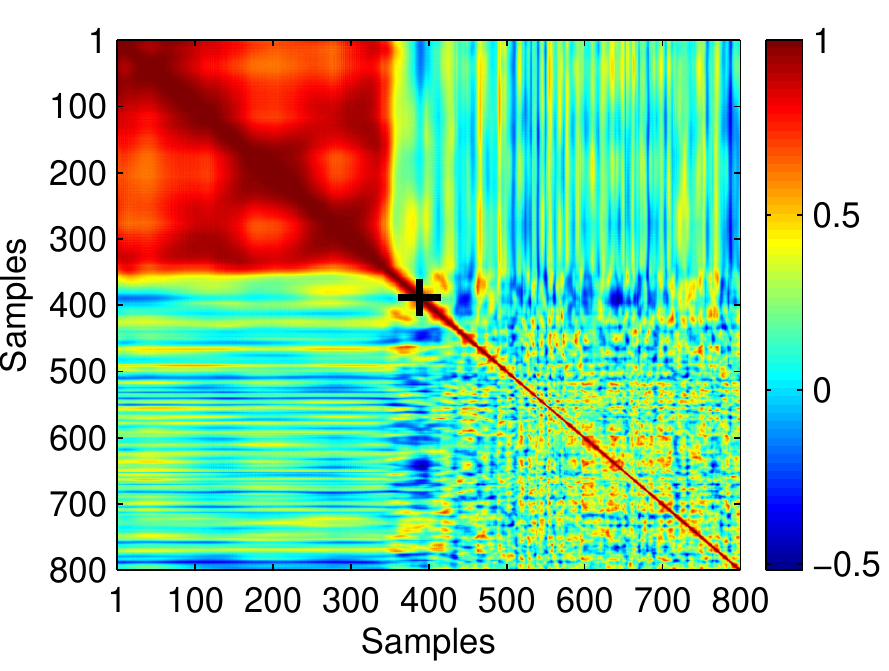}\\
(a) CSI View & (b) Correlation Coefficient\\
\end{tabular}
\end{center}
\caption { (a)  the CSI image of a sample. The movement starts at about 400
time-slot. (b) the correlation matrix of the CSI images shown in (a), the black
cross denotes the detected start. }
\label{Fig:motionDetection}
\end{figure}

Fig.~\ref{Fig:motionDetection} (a) presents a CSI sample, where the user
starts moving around 400th sample.  Observing the degree of disorder before
and after the start, we devise a correlation-based start/stop detection method.
The basis on the method is to leverage the rapid spatial de-correlation property
of CSI.  We find that the correlation coefficient $\rho$ between consecutive CSI
samples will drop rapidly if the spatial distance $d_s$ between them
is larger than  $\lambda/2$. Thus, there will be a rapid
co-efficiency raising or drop to check the ``moving'' and ``static''
status. Fig.~\ref{Fig:motionDetection} (b) shows the samples'
correlation matrix.
When device is static, stable and high correlation co-efficiency holds
the entire upper-left area, while it disappears immediately when the
device starts moving. According to our
experimental evaluation, a devices is said to be moving when $\rho$
drops below $0.4$, and the final detected time $t_d$ is quite close to
the actual time $t_a$.

\subsection{Minimum Sampling Rate}
\begin{figure}[hptb]
\begin{center}
\includegraphics[scale=0.6]{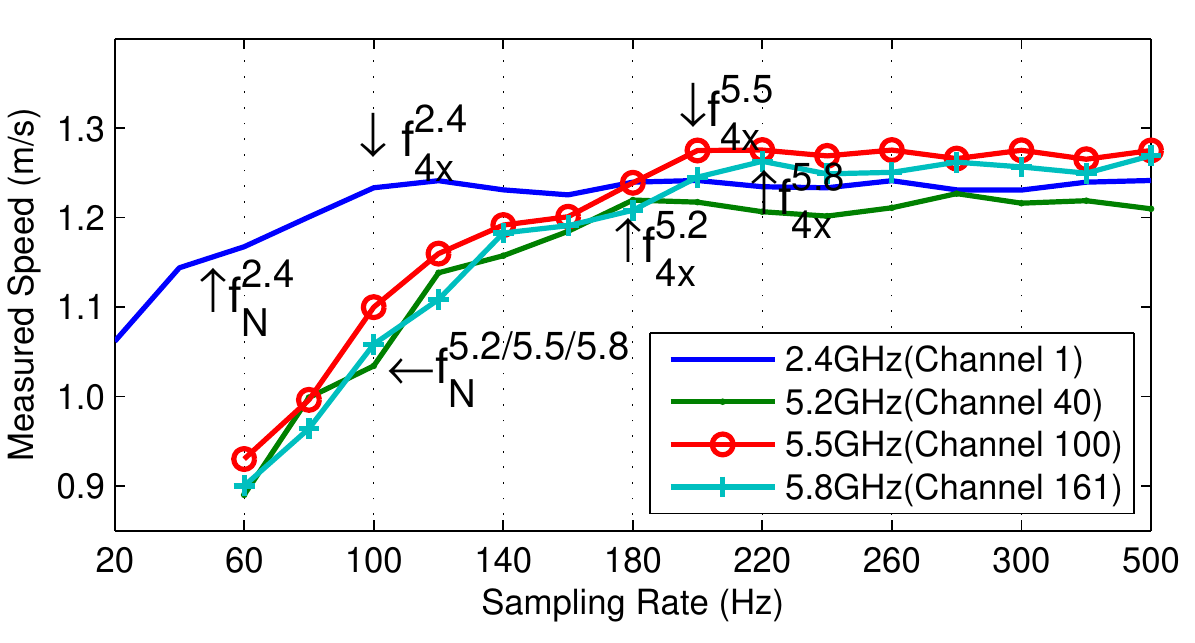}
\end{center}
\vspace{-0.3in}
\caption{The measured speeds in different sampling rate.  The points
  denoted by $f^\star_N$ denotes the Nyquist sampling rate, while
  $f^\star_{4x}$ denotes the minimum require sampling rate.}
\label{fig:sampling}
\vspace{-0.1in}
\end{figure}

Similar to sensor-based system, sufficient CSI sampling rate is critical for accuracy.
Due to the un-equal time distribution between fading and non-fading,
 the Nyquist sampling rate of $f_s=2\cdot f_{max}$ is not
 sufficient.
We carried out experiments to find the minimum $f_s$ that
 can guarantee good accuracy.
Evaluations are carried out in a wide range of channel frequencies
including 2.4G (channel 1), 5.2Ghz (channel 40), 5.5G (channel 100)
and the highest 5.8G(Channel 161).
During the experiment, testers are walking at the same speed around
 $v=1.3m/s$ and the mobile device in their hands are constantly
 transmitting beacon frames at 500hz.
After the experiments, we simulate the sampling rate $f_s$ from 20hz
 to 500hz by dropping frames uniformly.
 Fig.~\ref{fig:sampling} presents the results. We can see from the figure that
 the estimated speed $v$ continuously climbs when $f_s$ is higher than
 Nyquist rate $f_N$, and the speed stops raising when $f_s$ is about 4
 times of $f_{v}=2v/\lambda$. More experiments in other situations have
 also confirmed the $4\cdot f_v$ sampling rate. Therefore if we set
  $f_{max}=1.6m/s$, the minimum sampling rate is only 100 (or 250)
 frames/s in 2.4G (or 5.8G) environment,  or equivalent to approximately
 40KBps or 100KBps traffic.

\begin{figure}[hptb]
\begin{center}
\includegraphics[scale=0.6]{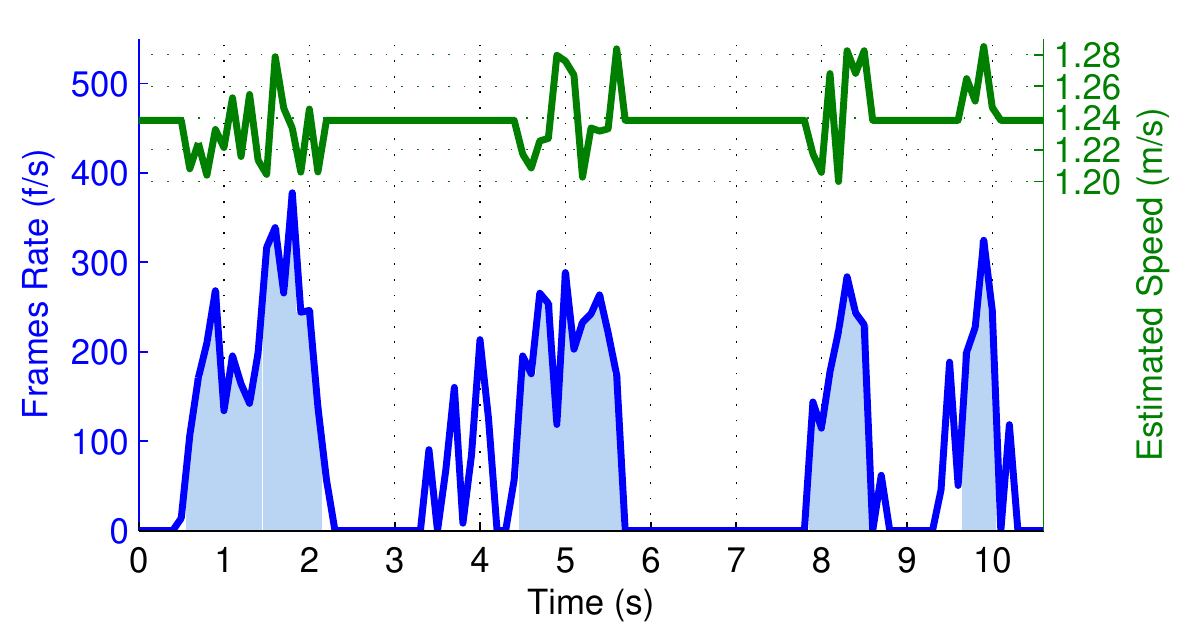}
\end{center}
\vspace{-0.3in}
\caption{The burstiness of wireless traffic in practical environment
  and the estimated speed. }
\label{fig:burstiness}
\vspace{-0.2in}
\end{figure}

The traffic burstiness is another problem. The burstiness, which
happened frequently, is obviously against the CSI-based speed
estimation. Since the burstiness is usually short-time high-frequency
traffic phenomenon, a reasonable assumption could be made to ease this
problem: people's walking speed remain stable during the gap between
two burstiness. Fig.~\ref{fig:burstiness} presents our solution that
during each burstiness the speed is estimated, while in the gap the
speed is approximated as the average.

\section{RSS Fingerprints Extraction}
\label{sec:fingerprintExtraction}
RSS fingerprint are the most representative RSS point extracted from many
samples for a given
 position, and the error of the fingerprint will directly affect the mapping and
localization accuracy.
 Besides the strong noise in RSS measurement, there are two main type of errors.
 The first is the well-known device-based measurement offset, which is mainly
caused by the variation of device antenna. We observe another measurement error,
called \emph{directional shadowing problem}, which has minor effect on
traditional site-survey but has strong interference on crowdsensing based
approach.

\subsection{The Directional Shadowing Problem}
\begin{figure}[t]
\begin{center}
\includegraphics[scale=0.8]{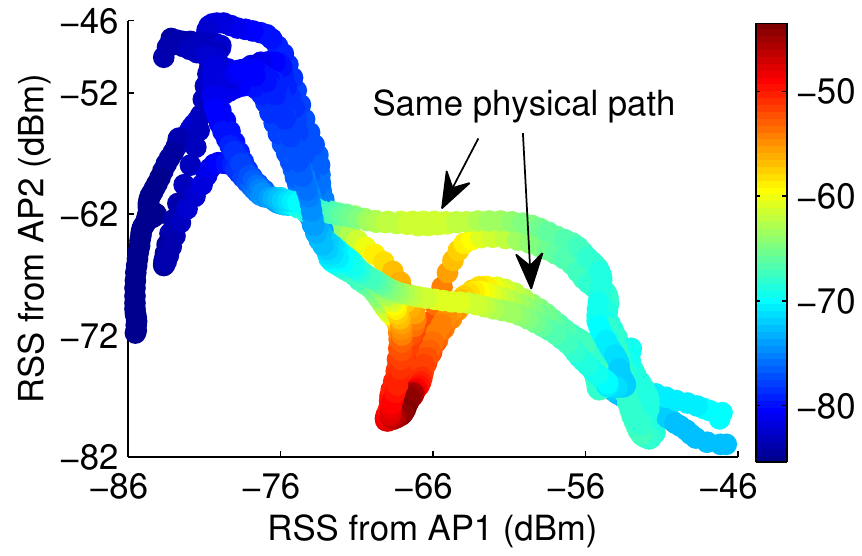}
\end{center}
\vspace{-0.1in}
\caption{4 RSS trajectories measured from 3 APs. The color denotes the value for 3rd AP:
deeper color denotes higher  RSS for the 3rd AP. }
\vspace{-0.2in}
\label{fig:rssdifference}
\end{figure}

Human body is a strong electromagnetic energy absorption object, which may cause
directional shadowing effect to wireless signal. Therefore, when the same device
is placed at different position near human body, \eg shirt pocket or back
pocket, the RSS measurements will be obviously deviated.

This effect has been exploited in previous works~\cite{zhang2011antenna,spinloc}
to achieve Direction-of-Arrival (DoA) detection, however, it may cause severe
error to unsupervised fingerprints-to-floor plan mapping, because there may be
multiple parallel RSS trajectories corresponding to the same physical path.
Fig.\ref{fig:rssdifference} presents an example of directional shadowing from
real measurement data. In this situation, the mapping algorithm may wrongfully
think that there are parallel paths between the start and end, and it is very
likely to cause mapping failure.

\subsection{Fingerprints Extraction}

Previous approaches usually adopted cluster-based algorithm to extract
fingerprints by merging the the nearby RSS samples within a certain threshold.
This is a coarse-grained algorithm, and it cannot identify the device offset and
directional shadowing. Our solution comes from a intuitive observation: although
the RSS samples are highly deviated for the same physical position if under
antenna variation and directional shadowing, the temporal RSS samples transition
trends are very similar. If we could correctly identify the matching between
these RSS trajectories, the fingerprints could be extracted without two errors
mentioned above.

This intuitive idea could be transformed to a a $n$-partite graph matching
problem if we see these trajectories as curve-shaped graphs.  This can further
be done by $n-1$ times iteratively graph matching between $n$-th trajectory and
previous resulting RSS fingerprints graph. As revisited in Section
\ref{sec:related}, relaxation-based approaches approximate the graph matching
elegantly, and the only job is to build the affinity matrix $\mathcal{M}$.

Given two RSS trajectories, $G^P$ and $G^Q$, each of them contains $n^P$ and
$n^Q$ RSS measurements respectively. A possible assignment $a$ is defined as
$a=(i,i')$ , where $i\in \mathcal{V}^P$ and $i' \in \mathcal{V}^Q$. Given
another possible assignment $b=(j,j')$, the compatibility $m_{a,b} \in
\mathcal{M}$ is assigned as follows in current system.
\begin{equation}
\mathcal{M}_{(a,b)}= \left\{ \begin{array}{rl}
e^{-||d_{ij}-d_{i'j'}||}, &\mbox{ if $||d_{ij}-d_{i'j'}||>\epsilon$ } \\
0, &\mbox{ otherwise} \\
 \end{array} \right.
 \label{eq:m}
\end{equation}
where $d_{ij}$ and $d_{i'j'}$ are the \emph{distance}
between $i$ and $j$, and their assignment pairs $i'$ and $j'$,
respectively. Since RSS attenuation along distance is non-linear, in
our system the Minkowski distance \cite{minkowski} with value $p=1.7$
is used to define the distance between RSS samples $a$ and $b$.
 \begin{equation}
d_{ab} = (\sum^n_{i=1}{|rss_{ai}-rss_{bi}|}^p)^{\frac{1}{p}}
\end{equation}
where $n$ represents the numbers of all heard APs, and $rss_{ai}$ for the $i$-th
AP's RSS value of RSS sample $a$.

\begin{figure}[t]
\begin{center}
\includegraphics[scale=0.8]{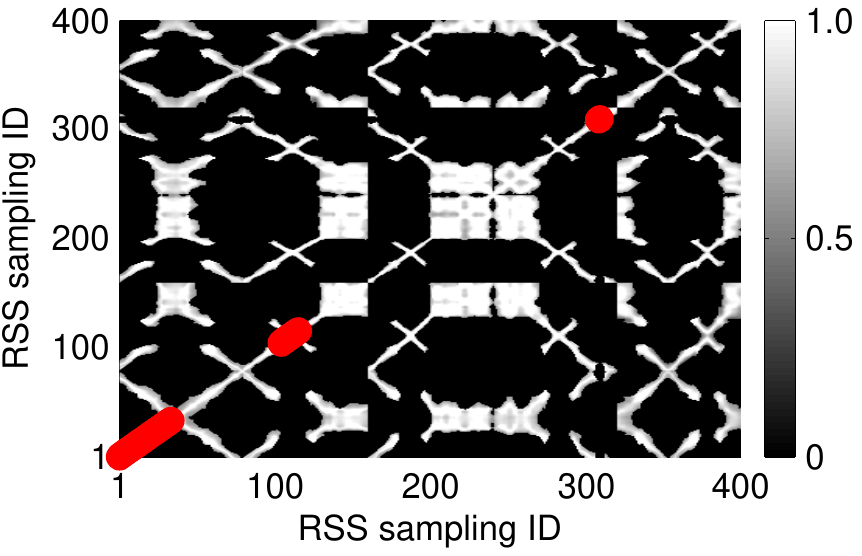}
\vspace{-0.1in}
\caption{The correspondence ratio among RSS sampling sequence.
  Ted dots denote the extracted RSS fingerprints.  }
  \vspace{-0.2in}
\label{fig:matrixAndMDS}
\end{center}
\end{figure}

In each iteration the fingerprints extraction is a partial graph matching
problem, and it is not tight integer constrained. We use spectral matching (SM)
algorithm to calculate the optimal column-wise vector $\mathcal{X}$ with length
$n^P \times n^Q$. $\mathcal{X}$ is further reshaped to an \textit{association
matrix} $\boldsymbol{A}^{P  \times Q}$, where each element $\boldsymbol{A}_{ij}$
denotes the \textit{matching rate} between $i \in \mathcal{V}^P$ and $j \in
\mathcal{V}^Q$. Fig.~\ref{fig:matrixAndMDS} presents an example of the
association matrix between two identical RSS trajectories, where high value
denotes highly possibility of matching.
\subsection{Fingerprints Transition Graph}
The Fingerprints Transition Graph $\mathcal{G}^F=(\mathcal{V}^F, \mathcal{E}^F)$
records the spatial connectivity of all fingerprints.  Since the vertices set
$\mathcal{V}^F$ is the fingerprints set $\mathcal{C}^F$, we only need to
determine the edge set  $\mathcal{E}^F$ and the weight of edges
$W(\mathcal{E}^F)$.

Basically, any two nodes $i,j\in \mathcal{V}^F$ will an edge $e_{ij}\in E^F$ if
they satisfy following two conditions:

\begin{enumerate}
\item $i$ and $j$ are
subsequent RSS fingerprints within the same RSS fingerprints trajectory;
\item or when $i$ and $j$ belong to different trajectory, the distance $d_{ij}$
is smaller than $\epsilon$ and at least one of both is the $start$ or $end$ of a
trajectory.
\end{enumerate}

The weight of edges will be set to the absolute distance between fingerprints in
traditional approaches~\cite{ez, lifs, Zee}, however, in \ourprotocol, there is
no absolute distance information due to the absence of physical space
measurement.

Fortunately, the walking duration becomes a distance indicator, and we can
assign \emph{virtual distance} to the weight of edges.
The virtual distance is based on a reasonable assumption: although people may
vary their walking speed according to mood or ongoing tasks, they usually
maintain a constant speed during a single walking trip.
Based on this assumption, we can infer the distance ratio between nearby
fingerprints sequences along a trajectory, and eventually obtain a global
distance matrix.
Since the assignment of virtual distance for a fingerprint sequence is gradually
propagated, error accumulation may exist. However, the logical isolation of
indoor space is very small, the virtual distance for most of segments can be
assigned in less than 3 iteration, the error propagation is therefore ignorable.
Our virtual distance assignment algorithm is briefly described as blow.

\section{Mapping Between Fingerprints and Floor Plan}
\label{sec:crowdsourcing}
Ordinary floor plan is not friendly for crowdsensing based approaches,
especially when client-side doesn't provide direction information. By studying
the shortest walking distance (SWD) in indoor space, we realize that there is a
highly curly 2D manifold embedded in a 2D floor plan polygon.  The unfolded
version of this manifold will remove the direction information but preserving
the SWD information, which is friendly for crowdsensing-based approaches.

The technical underpinning of automatic establishing the mapping is based on the
topological similarity between the unfolded version of floor plan and the
fingerprint transition graph, which is a typical graph matching problem.
However, challenges exist in many aspects, \eg unacceptable computation cost and
low accuracy of large-scale GM, scaling effect, and the unsupervised nature of
auto mapping. Here we start from transforming the floor plan, then go through
the auto mapping algorithm.

\subsection{Floor Plan in Manifold's Eyes}

$n$-dimensional manifold is a topological space that near each point resembles
$n$-dimensional euclidean space, while globally not euclidean.
The indoor floor plan shares the same property.
Due to the obstruction of walls, the shortest walking distance (SWD) between two
points $i$ and $j$
in the floor plan $\mathbf{P}$ equals to the euclidean distance $d_{eu}(i,j)$
$\mathbf{iff}$ the points $i$ and $j$ are within the same local isolation with
direct Line-of-sight distance. When they are not in the same isolation, the SWD
would be the geodesic distance $d_{geo}(i,j)$ which detour through various of
obstructions.
 In this way, indoor space could be essentially viewed as a 2D-manifold
$\mathfrak{S}$ embedded
in a 2D-polygon $\mathbf{P}$.

In a manifold space, the euclidean distance is misleading. The geodesic distance
actually reveals the true structure of the manifold. Therefore, we resample the
2D floor plan  $\mathbf{P}$ using $n$ points uniformly scatted. A $n$ points
graph $G^M$ is then constructed. For every pair of nodes $i,j \in V^M$, there is
an edge $e_{ij}$ \textbf{iff} the correspondence points $i^P,j^P$ in the floor
plan $\mathbf{P}$ are in their mutual neighborhood with \textit{direct
line-of-sight} distance, and the weight for edge is the direct distance that
$W(e_{ij})=d_{eu}(i^P,j^P)$.
matrix $A^M$, which records the pairwise geodesic distance between all sampling
points.

\subsection{Unsupervised Accurate Mapping}
Since the RSS samples are measured along users' walking trajectories, the RSS
fingerprints transition graph $G^F$
also share the same floor plan manifold structure.
 An intuitive idea of establishing the fingerprint map is to apply graph
matching algorithm directly
upon $G^{F}$ and $G^M$. However, the accuracy and performance of
large-scale graph matching ($>$50 points) is very poor for unsupervised
situation.
A lightweight relaxation to the problem is to apply graphing matching only on
corridor points. Once the corridor points graph are mapped correctly, it is easy
to match the rooms points.
Unfortunately, the accuracy and performance of graph matching between such
corridor points graphs is still not satisfied.

\begin{figure}[t]
\begin{center}
\includegraphics[scale=0.93]{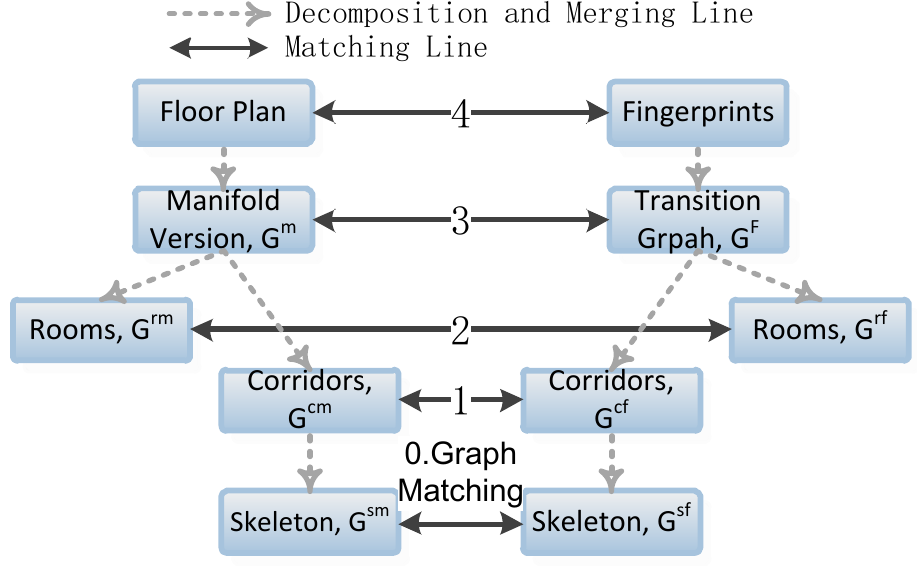}
\end{center}
\caption{The Skeleton-based matching algorithm. Skeleton graphs are first
extracted from both side.  Based on the matching between skeletons, upper levels
are gradually matched.}
\vspace{-0.15in}
\label{fig:mappingflow}
\end{figure}

Fortunately, the highly sparse chain structure of corridor points give us a
hint, and we devise a method called ``Skeleton-Based Matching'' to achieve
unsupervised accurate mapping between $G^{F}$ and $G^M$ even for very complex
indoor environment.
The basic idea is that: due to the high sparsity and chained structure of
corridor points graph, we can extract a coarse-grained skeleton graph from it
while preserving the identical topological structure. Adopting GM algorithm on
skeleton graphs will result high accuracy and performance.  Once the main
structures of $G^{F}$ and $G^M$ are matched, the rest parts will be matched
easily. Fig.~\ref{fig:mappingflow} illustrates the basic idea of skeleton-based
matching.

The algorithm is detailed in following 3 steps, including \emph{skeleton graphs
extraction}, \emph{skeleton graphs normalization}, \emph{skeleton graphs
matching}, and \emph{find-grained points matching}.

\subsubsection{Extract the skeleton graph}

Two sub-steps are required to extract the skeleton graphs.
1). Identifying the corridor points graph $G^{CF} \in G^F$ and
$G^{CM} \in G^M$;
2). extracting skeleton graphs $G^{SF}$ and $G^{SM}$ based on corridor points
graph $G^{CF} \in G^F$ .

In the first sub-step, a customized centrality measure $C(V)$ is
 designed to identify the core corridor network.
For a given point $v\in V$, its centrality $C(v)$ is measured as follows.
\begin{equation}
C(v) = \sum_{s\neq v\neq t\in V}\sigma_{st}(v)
\label{eq:centrality}
\end{equation}
where $\sigma_{st}(v)$ is the numbers of shortest path from $s$ to $t$ via $v$.

Based on this definition, we design a iterative algorithm to remove the
non-central
points effectively.
In each round of iteration, the centrality $C(v)$ is measured for every points.
If $C(v)$ is smaller than a low-bound $\tau$, then remove the points from the
graph. This procedure repeats until no points is removed.
Algorithm.\ref{alg:extractingCorridorPoints} shows the pseudo-code of corridor
points extraction.

\begin{algorithm}[!t]
\caption{Extracting Corridor Points}
\label{alg:extractingCorridorPoints}
\DontPrintSemicolon
\KwData{a sparse graph $G=(V,E)$}
\KwResult{a skeleton graph $G^S$ for $G$}

$G^T \gets G$, $done \gets$ false\;
\While{$done = $\textbf{false}} {
calculate centrality for $v\in V^T$, $done \gets$ \textbf{true} \;
\ForAll{$v_i\in V^T$} {
\If{$C(v_i)\leq \tau$}{
remove $v_i$ from $G^T$, $done \gets$ \textbf{false} \;
}
}
}
\end{algorithm}

\begin{figure}[b]
\begin{center}
\includegraphics[scale=0.66]{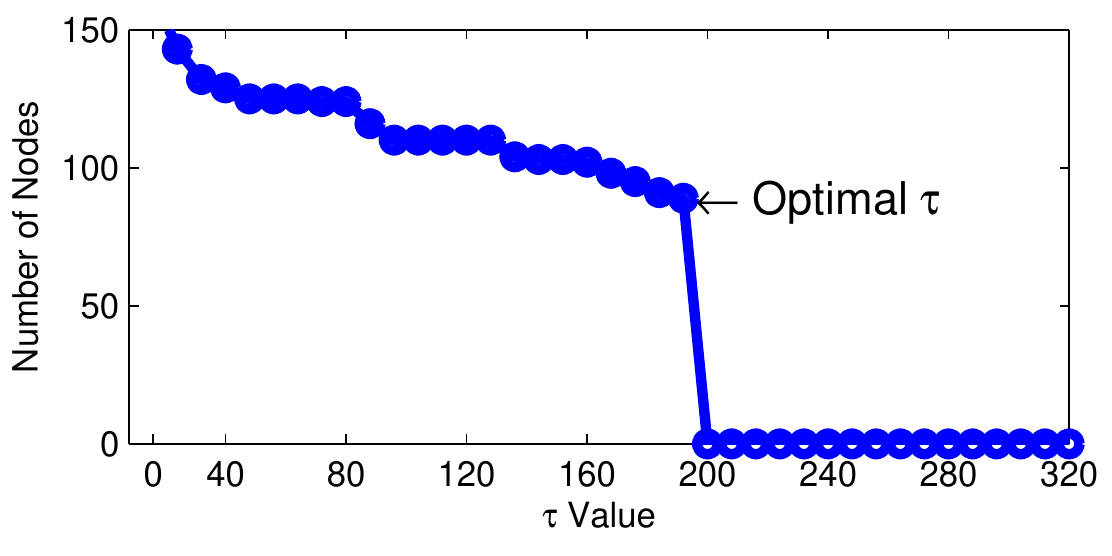}
\end{center}
\vspace{-0.2in}
\caption{Figure shows the linear growth of $\tau$, monotonic decrease of graph
size, and the optimal $\tau$ point.}
\vspace{-0.2in}
\label{fig:tauThreshold}
\end{figure}

The choose of value $\tau$ is critical. Fig.\ref{fig:tauThreshold} shows the
relationship between $\tau$ and size of graph. Small $\tau$ will bring in
redundant points, while large $\tau$ will remove all points. Since the linear
growth of $\tau$ will lead to monotonic decrease of number of remaining points,
we use binary search to find the optimal $\tau$ point.

In the second sub-step, the skeleton graphs $V^{S}$ is generated by
clustering the corridor points graph $G^{CF}$ and $G^{CM}$. We use
spectral clustering~\cite{von2007tutorial} (SC) as the clustering
algorithm.
SC is computationally faster than K-means and it only requires the
adjacency matrix which is exactly suitable in our case that both
$G^{CF}$ and $G^{CM}$ are represented only in adjacency matrix.
By clustering on $G^{CF}$ and $G^{CM}$, we obtain the vertices set of
skeleton graph $G^{SF}$ and $G^{SM}$.
The edge set $E^{SF}$ and $E^{SM}$ follow the underlying
points, that if two points $i,j$, belonging to different clusters$c_a$
and $c_b$ respectively, have an edge, then there is an edge between
$c_a$ and $c_b$. The weight of edge $e_{ab}$ is defined as the
shortest distance between the \emph{central points} of cluster $a$ and $b$, and
the
\emph{central point} of a cluster is the point $i$ which has the
shortest distances to other points within the cluster.

\subsubsection{Graphs Normalization}
Since the graph matching algorithm assigns the matching score according to
pairwise distance,  $G^F$ and $G^M$ must be normalized to same scale so as to
guarantee an accurate matching result.

The scaling effect can be undone by some global normalization of the graph
shape. Possible ways includes normalization of bounding box, or normalized
Laplace-Beltrami eigenvalues (LBE). The bounding box approaches works only for
rigid graph, while LBE is sensitive to deformations.
We use the commonly accepted  \emph{longest geodesic distance} $L_{lg}$ as scale
indicator.

edge weights of $G^{SF}$ and $G^{SM}$ are divided by the scale indicator. In
this way, the scaling of both graph is canceled.

\subsubsection{Skeletons Matching}
After the extraction and normalization of skeleton graphs,
we now find the best correspondence between $G^{SF}$ and $G^{SM}$. Let positive
symmetrical square matrix $M^{SR}$ and $M^{SM}$  represent their adjacency
matrices.
We build the affinity matrix $M^{{{SR}\times {SM}}}$ for graph matching as
follows.
\begin{equation}
M^{{{SR}\times {SM}}} = e^{{(\boldsymbol{1}^{SM}\bigotimes M^{SR} -
    \boldsymbol{1}^{SR}\bigotimes M^{SM})}^2}
    \label{eq:skeletonmatching}
\end{equation}
where $\bigotimes$ denotes the Kronecker product~\cite{kronecker} and
$\boldsymbol{1}^{SR}$ denotes the full-$\mathbf{1}$ matrix with the same size of
$G^{SR}$.
The idea behind Eq.~\ref{eq:skeletonmatching} is to enumerate all possible
matching candidates and store them in a large adjacency matrix $M^{{{SR}\times
{SM}}}$.
We use RRWM algorithm~\cite{cho2010reweighted} to perform the graph matching,
and Hungarian algorithm is further applied to discretize the $\mathcal{X}$ in
order to meet the final integer constraints $\mathcal{X} \in \{0,1\}^n$.

\subsubsection{Corridor Points Matching}
Although $G^{SF}$ and $G^{SM}$ are matching in previous step, however, the
corresponding clustering groups in $G^{CF}$ and $G^{CM}$ are not necessarily
matched due to the inconsistency of clustering operation.

In the corridor points graphs, we notice that only a few points connect multiple
chain structures, and they may serve as the \emph{bridge points}. Since the
topological structures of $G^{CF}$ and $G^{CM}$ are identical, if we could
identify the correct correspondences of these \emph{bridge points}, the points
within the chain structure will be matched easily.
In order to identify the bridge node, we introduce a new metric called
''bridge centrality'', which is equal to the number of shortest paths
 from all vertices to all others within nearby clusters that pass
 through that node.
\begin{equation}
C_{bg}(v) = \sum_{\{(s,t)|v\in c^i, s,t\in NN(c^i), s,t\notin c^i\}}
\sigma_{st}(v)
\end{equation}
where $NN(c^i)$ denotes the nearby clusters around $c^i$. The bridge point
will be the point with the highest bridge centrality.

After identifying the of correspondence of bridge points in $G^{CF}$ and
$G^{CM}$,  the chain structures are easily matched according to the start and
end bridge points.

\subsubsection{Rooms Points Matching}
If we remove the corridor points graph $G^{CM}$ from $G^{M}$, the rooms points
will naturally forms several clusters.  Each room points cluster $C_R$ connects
to the corridor by a door point, and the rooms matching is also easy by matching
the door points in both $G^{F}$ and $G^{M}$.

But if there is more than one rooms connect to a single door point in the
corridor, \eg two rooms in opposite side along the corridor, there may be
mis-matching. This kind of mismatching can be canceled using coarse-grained
propagation model.
Along with the corridor points matching, the coarse-grained position estimation
for APs can be done. With the rough location of APs, the rooms mis-matching can
be easily eliminated by checking the RSS values. The smaller RSS difference
means higher probability of being in the candidate rooms.

$\mathbf{P}$, the extracted RSS FTM also reveals the true structure the
manifold, which means the RSS FTM graph $G^T$ and $\mathfrak{S}$'s unfolded
version $G^M$ are naturally matching, and we are here to find the matching.b
$i\in V^P$ and $i' \in V^Q$. We next define the \textit{affinity matrix} $M$
that $M(a,b)=f(i,i',j,j')=f(a^P_e,a^Q_e)$, where $a^P_e=A^P(e_{ij})$ and
$a^Q_e=A^Q(e_{i'j'})$, and $f(\cdot)$ measures the compatibility of assignments
$a$ and $b$. Higher $M(a,b)$ denotes higher compatibility and vice versa. In
such setting, the graph matching problem transformed to finding the optimal set
$C$ of assignments that will maximize the inter-cluster score $S=\sum_{a,b\in
C}M(a,b)$. Cluster $C$ can be further represented by an indicator vector $x$,
and we can rewrite the inter-cluster score as:
$x^TMx$
inter-cluster score $S$ is the principal eigenvector of $M$, and there are
various methods to find the principal eigenvector $x^*$.
%
$M$. In our setting, $M$ is a $n^P\times n^Q$ symmetric non-negative matrix,
where each element $M(a,b)$ is defined as

%
the points distance.
manifold is not globally equal to the euclidean distance that $\forall
p_i,p_j\in \mathfrak{S}$: $d_{eu}(p_i,p_j)\leq d_{geo}(p_i,p_j)$,
smaller than the euclidean distance.
%
distance between RSS fingerprints along users' walking trajectories. Apparently,
the original indoor space manifold $\mathfrak{S}$ should be transformed to an
unfolded version so as to help establish the mapping between the RSS space and
indoor space.
%
to transform the $\mathfrak{S}$ into a Manifold Unfolded Space (MUS). In MUS the
euclidean distance between any two points is equal to the geodesic distance on
original manifold $\mathfrak{S}$, $\forall p_i,p_j\in
S^p:d_{eu}^{S_{eu}}(p_i,p_j)=d_{geo}^{\mathfrak{S}}(p_i,p_j)$.
%
geodesic distance matrix $D_{geo}$ to the classical multi-dimensional scaling
(MDS) \cite{mds}, and matrix $D_{geo}$ is obtained by calculating the shortest
walking distance between all pairs of points on the manifold.
%
in Section.\ref{sec:fingerprintExtraction}, the correspondence matrix $X$ is
still obtained through the process of building compatibility matrix $M$ and
calculating the principal eigenvector $x*$. What is different from the
trajectories matching is the distance function for adjacency matrix of two
graphs, here the distance is the euclidean distance in MUS.
structure is of highly symmetry, the correspondence shown in $X$ may contain
multiple candidate mapping.
methods.
%
to obtain the coarse-grained RSS distribution in the indoor space. The reversion
of predicted RSS distribution at some sampling position will correspond to the
predicted RSS measurement of the APs.
the RSS distribution obeys the estimated trends. Therefore, the true
correspondence leaves behind.

\section{Localization and Tracking}
\label{sec:tracking}
\ourprotocol  provides a unified localization and tracking service by
treating the direction localization request as a tracking request
without historical data. Here we mainly focus on tracking technique in
\ourprotocol. Unlike the stateless K-NN based method which is widely
adopted in previous approaches, in our solution, the users'
trajectories are globally determined from the very beginning by
transforming the tracking problem to a graph matching problem between
the measured RSS samples transition graph $G^S$ and the fingerprint
transition graph $G^T$. After the graph matching, the accuracy is
further improved by bringing in the CSI-based speed estimation through
a particle-filter based  fusion. Here we start introducing these
two steps.

\subsection{Graph Matching Based Tracking}

Graph matching based tracking is to find the best correspondence
between the sequence of RSS samples  of tracking request and extracted
fingerprints. This is exactly the same matching process undertaken in
Section~\ref{sec:fingerprintExtraction}, except for the differences
that the tracking is to find the matched points, while in
Section~\ref{sec:fingerprintExtraction} the un-matched RSS samples
are added to the fingerprints database.

Let $\mathcal{X}^{n^S\times n^F}$ represent the association matrix
obtained through spectral matching where $n^S$ and $n^F$ are the
numbers of RSS samples and candidate fingerprints respectively. Due to
the error in RSS measurement and fingerprints map construction, a
single RSS samples $p_i$ may correspond to multiple fingerprints in
$\mathcal{X}$,\eg, a RSS sample may correspond to two fingerprints,
one is in corridor, and another is in a room. Fortunately the temporal
correlation can help eliminate those false correspondence by checking
the spatial continuity between current and subsequent candidate
fingerprints.
After eliminating the false correspondence, the globally estimated
coordinates sequence will be given by $T_{G^S} = f_{G^T}(G^S)$, where
$f:fingerprint \rightarrow G^M$ represents the mapping from
fingerprints to the floor plan manifold $G^M$, and $T_{G^S}$ are the
resulting coordinates sequence in $G^M$.

However, full graph matching between all fingerprints set and RSS
samples is time-consuming, therefore a  matching candidates pruning
process is introduced to meet practical demand of real-time tracking
of multiple clients.

\textbf{Matching Candidates Pruning}:
The main idea of the pruning is to find the probable walking area for
the tracking request using coarse-grained nearest neighbour (NN)
method, which will significantly reduce the search space.

%
Let $R^T=\{r^T_1,..,r^T_n\}$ represent the RSS sequence of the
tracking request at time $t$, where $r_i$ is a $n$-dimensional RSS
vector. The length $n$ is smaller than a positive integer
$\mathcal{L}$, such that the tracking algorithm provides limited
backtracking. Let $R^F=\{r^F_1,..,r^F_n\}$ denote the RSS
fingerprints set and graph $G^F_P$ denote the positions of
fingerprints in the floor plan. The pruned matching candidate set
$S_u$ is defined as follows.
\begin{equation}
S_u=\{r^F|r^F \in R^F, r^F\in\epsilon\text{-}\rm{NN}_{R^F}(NN_{R^F}(R^T))\}
\end{equation}
where $\rm{NN}_{R^F}(R^T)$ denotes the nearest neighbours of $R^T$
within the fingerprints set $R^F$, and
$\epsilon\text{-}\rm{NN}_{A}(B)$ denotes the $B$'s neighbourhood
within distance $\epsilon$ in the set $A$. Here the $\epsilon$ is set
to $3$m.

\subsection{Fine-Grained User Tracking}

The movement estimation from trajectories matching (TM) and CSI-based
speed estimation (CBSE) are naturally complementary. TM is
slow-responsive yet accurate in long time, while CBSE may drift with
time cumulation yet be accurate in short time. Obviously, a
fusion scheme for TM and CBSE will significantly improve the
accuracy. Due to the linearity constraints and difficulties in
correctly estimating the error covariance matrices, we don't use the
conventional Kalman filter. The more robust particle filter (PF) is
adopt as the fusion algorithm.

The state space of the tracking is a two dimensional vector
$X_t=[M_i,vt]$, where $M_i$ represents the $i$-th node in the floor
plan manifold $M$, and $v$ for the walking speed. The main challenge
here is that the speed $v$ has no direction information, therefore
every round the PF evolves and there are multiple candidate
predictions. For instance, when walking in a corridor without optional
paths, there are two candidate directions, forward and backward, and
therefore two state candidates $X_{for}=[M_j,v_t]$ or
$X_{rev}=[M_h,v_t]$, where $M_j$ and $M_h$ are two nearby nodes in
different directions.

 Fortunately,the PF can handle this problem elegantly by sprinkling
 different amounts of particles in different directions. In the
 prediction phase of each filter iteration, the particles will be
 re-sampled and enumerate all possible candidate directions.

\begin{figure}[htb]
\begin{center}
\begin{tabular}{cc}
\includegraphics[scale=0.45]{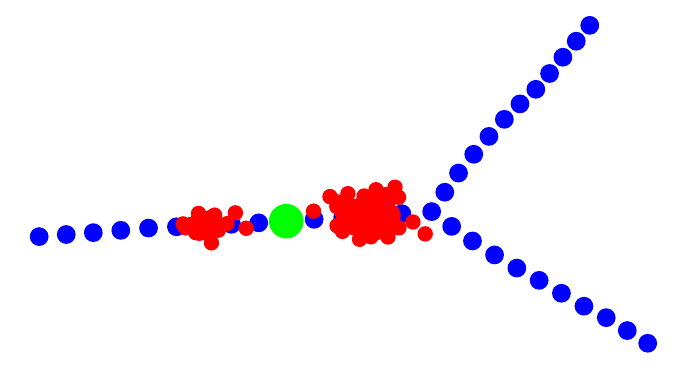} &
\includegraphics[scale=0.45]{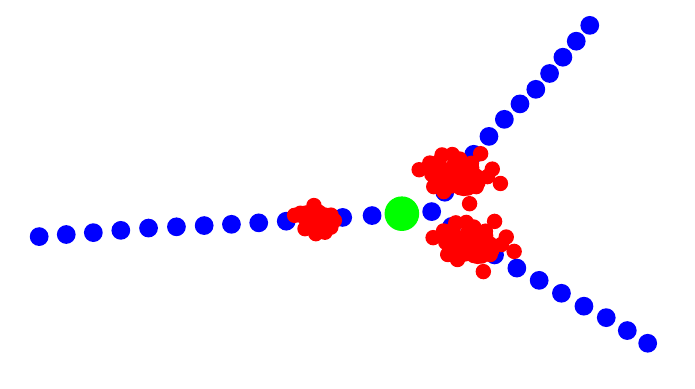}
\vspace{-0.1in}
\end{tabular}
\end{center}
\vspace{-0.2in}
\caption{The predicted particles distribution before each new
  measurement: red dots represent the particles, and green dots for the
  determined position in last round.  }
\label{fig:particles}
\vspace{-0.1in}
\end{figure}

 Fig.~\ref{fig:particles} illustrates the particles distribution along
 a path, in each round the particles enumerate all possible candidate
 directions. When there are optional paths, the particles will
  enumerate all possible options.
After the  fusion, the $M_i$ is transformed to floor plan
coordinates.

\section{System Evaluation}
\label{sec:evaluation}
\subsection{Prototype System with Customized APs}
\begin{figure}[t]
\begin{center}
\includegraphics[scale=0.5]{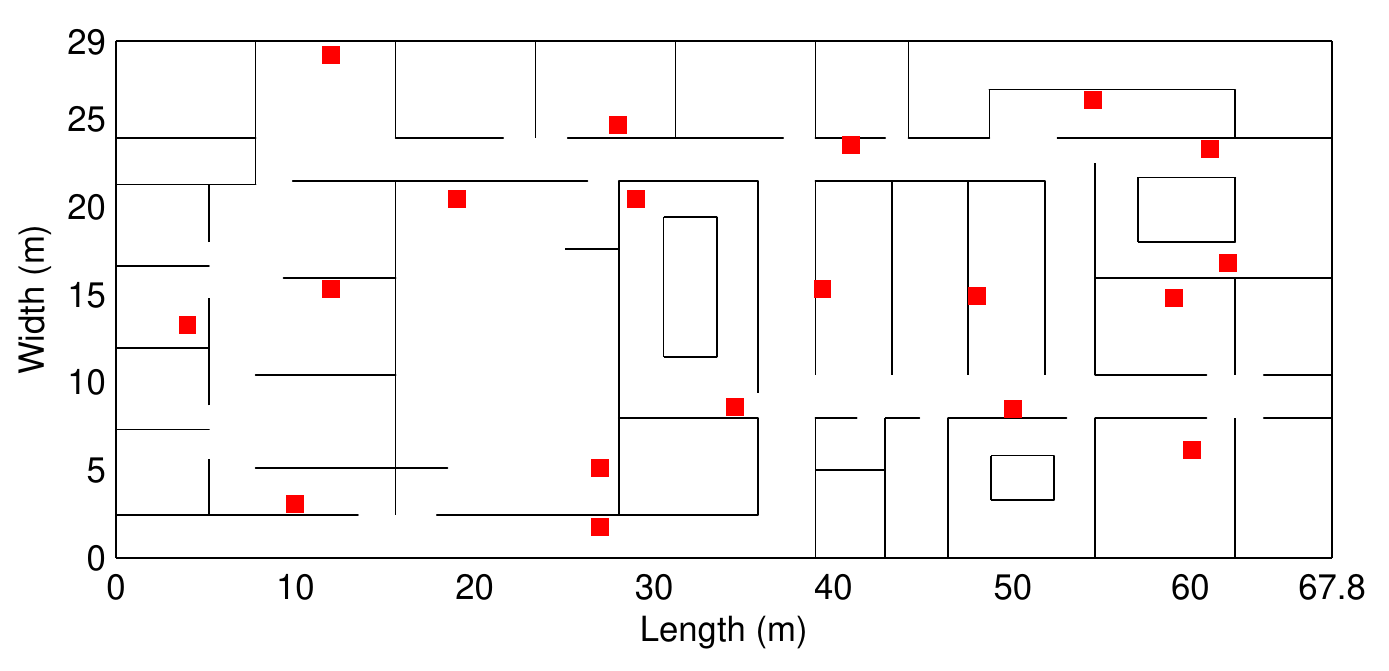}
\vspace{-0.1in}
\caption{The floor plan of our test area, and  red dots denote the
  deployed customized APs.}
\vspace{-0.1in}
\label{fig:qujiangmap}
\end{center}
\end{figure}

We designed and developed a prototype hardware system for \ourprotocol.
The  prototype system is deployed in a large $2000m^2$
office environment with circular corridor network as shown in
Fig.~\ref{fig:qujiangmap}. To maximize the participation rate,
\ourprotocol is deployed at AP-end.
In total, 18 customized APs are sparsely
 deployed across the office that provide both wireless networking
 service and localization service.
They will forward the measured RSS and CSI
 values to a central localization server.
A developed simple client App is installed on
 testers' Android smartphones.
The localization server will calculate the clients' position
 based on the networking traffic,
 and clients only need to read the location coordinate from server.

\begin{figure}[t]
\begin{center}
\includegraphics[scale=0.5]{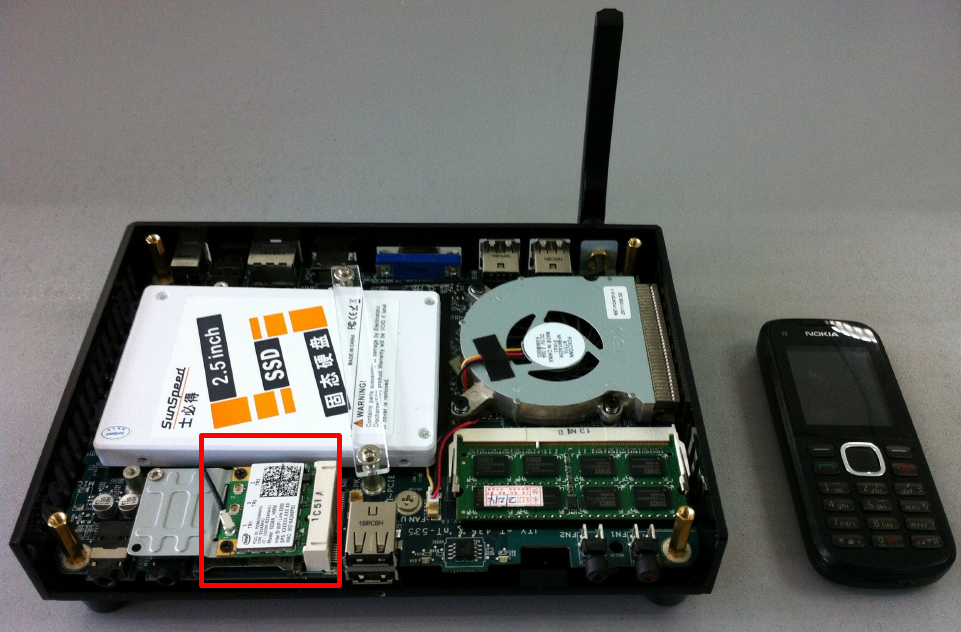}
\vspace{-0.1in}
\caption{Prototype system  for \ourprotocol with Intel Atom-based Mini
  PC and 5300NIC  as AP.}
\label{fig:appic}
\end{center}
\vspace{-0.2in}
\end{figure}

Although CSI is a standard PHY layer information, currently only
 Intel 5300 NIC can export it to user level. Our customized AP is
simply an Intel ATOM-based mini PC with 5300 NIC. Figure
\ref{fig:appic} shows our customized AP. It equips with single-core
1.6Ghz ATOM CPU and 5300 NIC. The total cost is about
\$90. The OS is Ubuntu 12.04 64-bit, and AP function is hosted by hostapd.
Besides the ordinary AP interface, a monitor virtual interface is also added to
overhear the wireless traffic. Both
measured CSI and RSS are uploaded to localization server in real-time.

\subsection{CSI-based Speed Estimation}
we made extensive experiments to evaluate the universality of CSI-based speed
estimation. Three groups of experiments mainly cover all wireless channel
combinations, propagation scenarios, and a comprehensive evaluation in a
typical office environment.
\subsubsection{Performance in different channel combinations}
In this section, we mainly focus on the performance of speed estimation in
different channels and transmission modulations (MCS). The experiments covered
all supported channels (channel 1-13 in 2.4Ghz, channel 38-163 in
5Ghz) and MCS values (MCS 0-7). During each $<$channel,MCS$>$ combination, a
testing laptop moves along a 42m long corridor, and it transmits random
messages at 800Hz in the given $<$channel,MCS$>$ combination. Several APs along
the path will record the CSI value during the movement.

\subsubsection{Performance in different propagation scenarios}
Performance in a typical strong multipath environment has been evaluated in
previous section. To evaluate the performance of speed estimation in a pure
non-multipath to moderate multipath environments, 3 groups of experiments are
carried out in a fully-open square, a rooftop platform with several cooling
infrastructures, and a large gymnasium. In each environment, 7 tests are
carried out to cover the most commonly used $<$channel,MCS$>$ combination,
$\left< { 6,0 } \right> $, $\left< { 40,0 } \right> $, $\left< { 60,0 }
\right>$, $\left< { 100,0 } \right> $, $\left< { 120,0 } \right> $, $\left<
{ 140,0 } \right> $, and $\left< { 161,0 } \right> $, covering the channels from
2.4Ghz to 5.8Ghz.

\subsubsection{Performance in practical office environment}
In this experiment, 10 students are asked to walk 3 times around the
$129m$ long circular corridor as shown in
Fig.~\ref{fig:qujiangmap}. They are asked to maintain constant walking
speed in first round, and the speed may change slightly and remarkably
in second and third round. The measured speed will be
integrated to walking distance $D_{w}$. We mainly considered the error
rate $e_w = \vert D_{path} - D_{w}\vert /D_{path}$.

Two other approaches, pedometer-based and channel coherent-time
based~\cite{speedestimation1}, are also developed  as comparison. For
pedometer we use NASC~\cite{Zee} method to detect steps, and the step
length is predefined according to the training data.  For
coherent-time based approaches, we use a constant $\xi=0.396$, which
is manually optimized specific for this experimental field.
The laptop held by students constantly transmits beacon-frames at
500Hz, and accelerometer data used for HD active protection is
recorded for step detection.
\begin{figure}
\begin{center}
\includegraphics[scale=0.5]{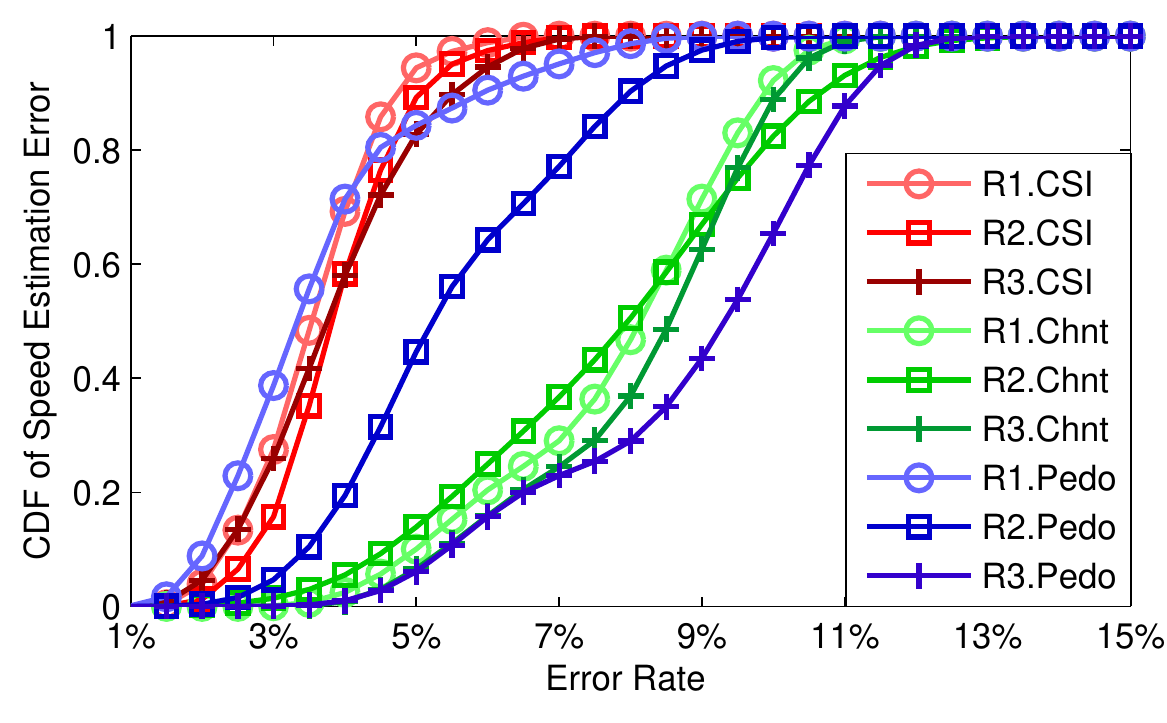}
\vspace{-0.2in}
\caption{The Error CDF of 3 methods.}
\label{fig:csiaccuracy}
\vspace{-0.2in}
\end{center}
\end{figure}

Fig.~\ref{fig:csiaccuracy} presents the error CDF of three methods in
 three walking manners: with constant speed
  (R1), slightly varying speed (R2), and with marked change (R3).
The  experiment shows  that,
 comparing to pedometer-based approach, \ourprotocol with CSI can
 achieve  better
 accuracy without requiring predefined constants or training data,
 meanwhile, \ourprotocol significantly outperforms pedometer-based
(Pedo) approaches  when speed is varying or pre-defined step length is
 out of  effectiveness.
Coherent-time based approach (Chnt) haves the similar
 speed-invariance feature, however, the accuracy is considerably poorer
 than \ourprotocol and pedometer in constant speed.

\begin{figure*}[tbh]
\begin{center}
\begin{tabular}{ccc}
\includegraphics[scale=0.6]{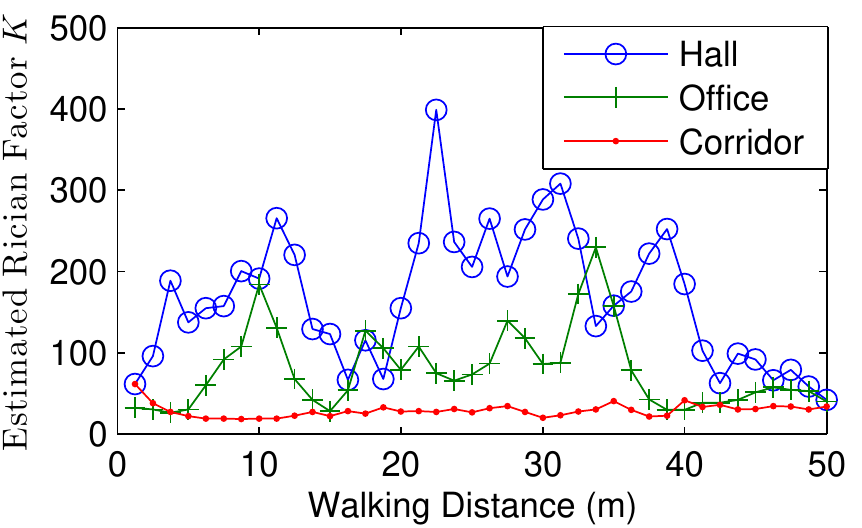}&
\includegraphics[scale=0.6]{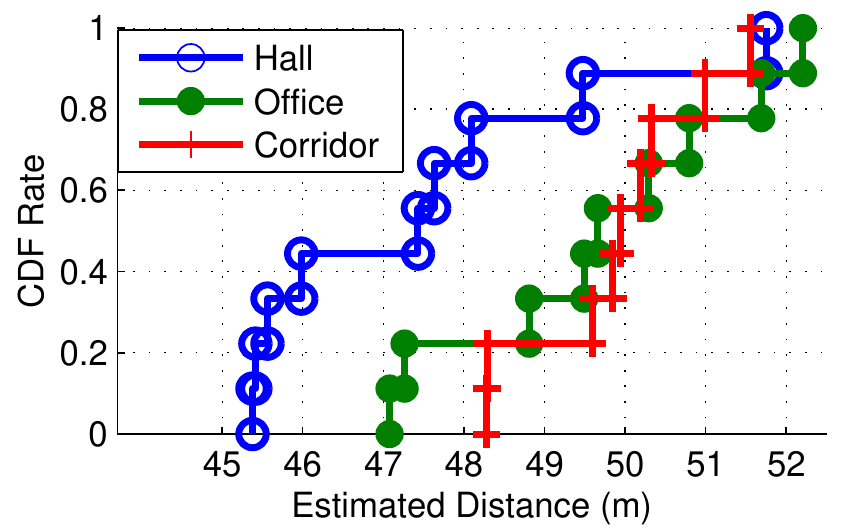}&
\includegraphics[scale=0.6]{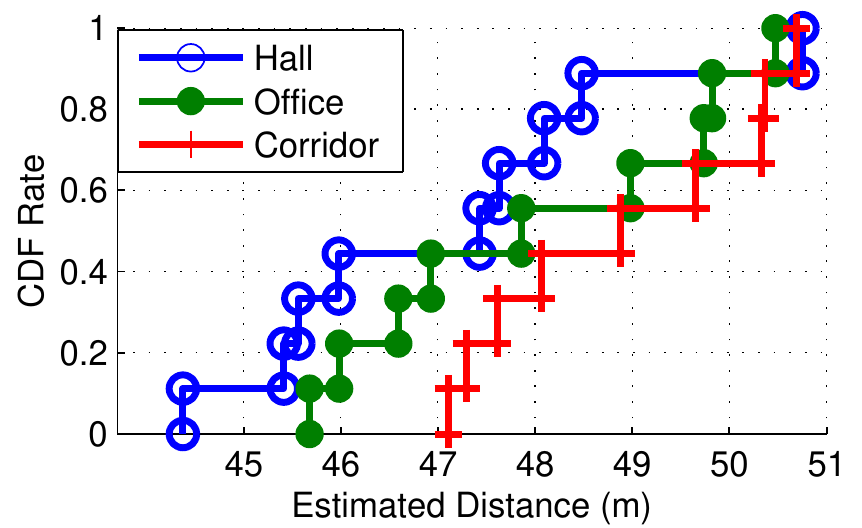} \\
(a) Racian Factor $K$ & (b)Strong Rayleigh Fading  & (c) Weak Rayleigh Fading
\end{tabular}
\end{center}
\vspace{-0.2in}
\caption{  (a)  the Rician factor $K$ along the path in different
  environment.  (b)  the CDF of estimated distance
  without LoS components  (c)  the  CDF   with strong LoS components.}
\label{fig:distance}
\vspace{-0.15in}
\end{figure*}

The accuracy of CSI-based distance estimation is then evaluated. To
evaluate the influence of richness of multipath components to the
distance estimation error, we carried out experiments in 3 typical
environments, a compact corridor, a large office environment, and a
very large hall. In each environment we walked along a 50m straight
line for 10 times. Two APs were simultaneously used to estimate
walking distance. One was placed at the end of the path with strong
LoS component, and the other was placed in a cubicle to cut-off the
LoS components to simulate strong Rayleigh fading.
Figure~\ref{fig:distance} (a) plots the Rician $K$
Factor~\cite{ricianKestimate} along the walking path
in different environment, which estimate the degree of LoS components.
 Very low and stable $K$ appears in corridor environment which
means there are rich multipath components, while in office and hall
the multipath components is significantly reduced due to the weak
reflection in large wide-open space. Figure \ref{fig:distance} (b) and
(c) plot the CDF of estimated walking distance by the AP in path and
cubicle respectively.
We see in the best situation that in a corridor with  strong
multipath component, there is only $3\%$ error.
In Figure~\ref{fig:distance} (c), we also see small
error happened  in the path, and in the worst situation,
in a large hall with very weak multipath components, the averaged
estimated error is less than $10\%$.

\subsection{Mapping Accuracy Test}

\begin{figure*}
\begin{center}
\begin{tabular}{ccc}
\hspace{-0.35in}
\includegraphics[scale=0.45]{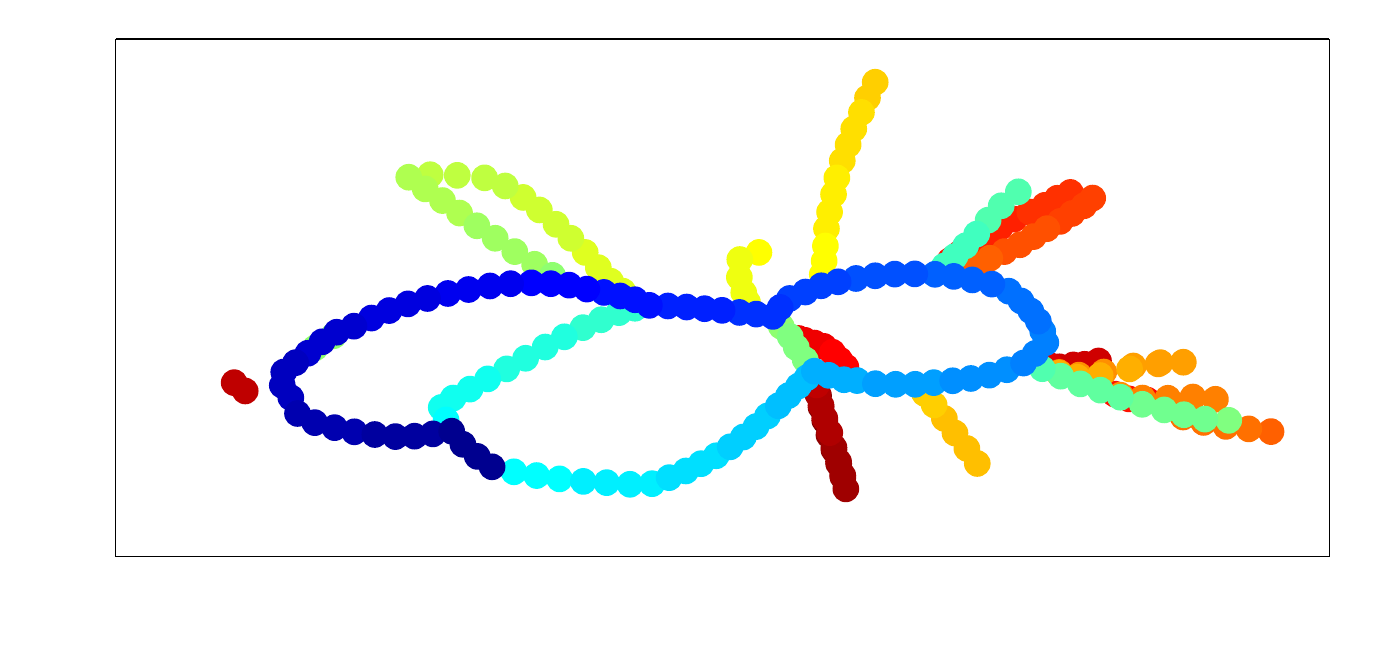}  &
\hspace{-0.35in}
\includegraphics[scale=0.45]{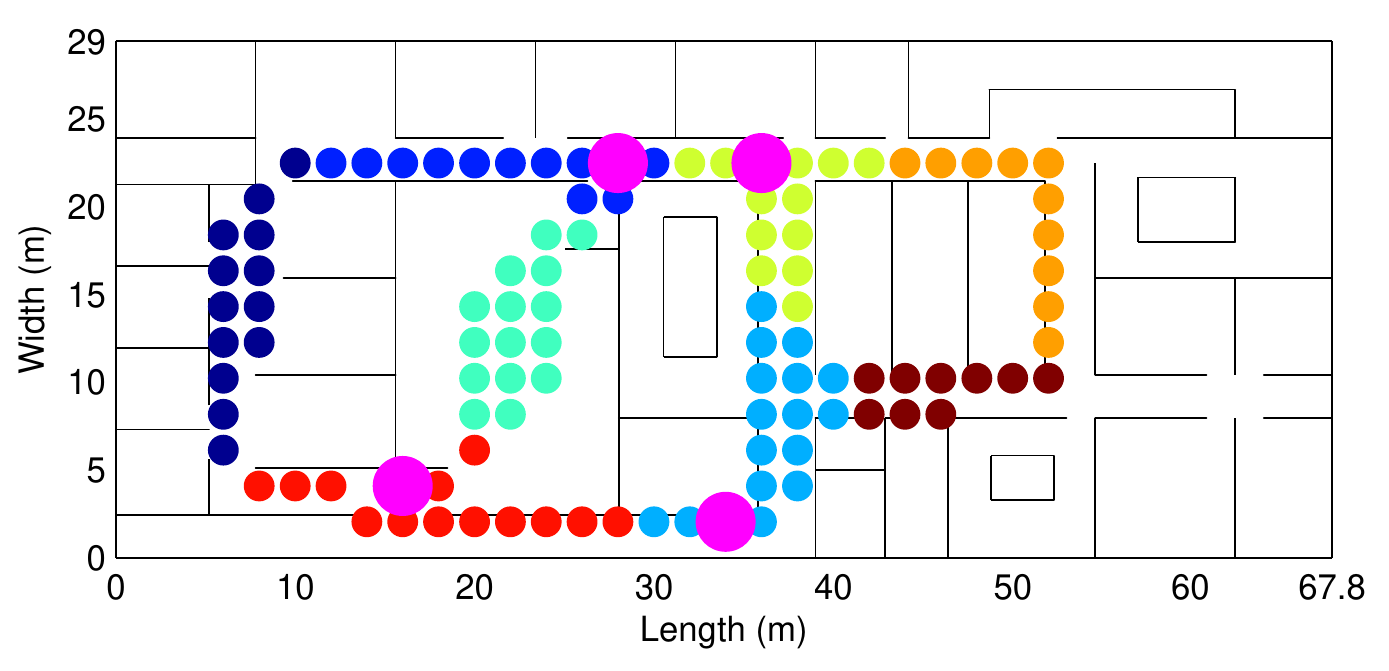} &
\hspace{-0.35in}
\includegraphics[scale=0.45]{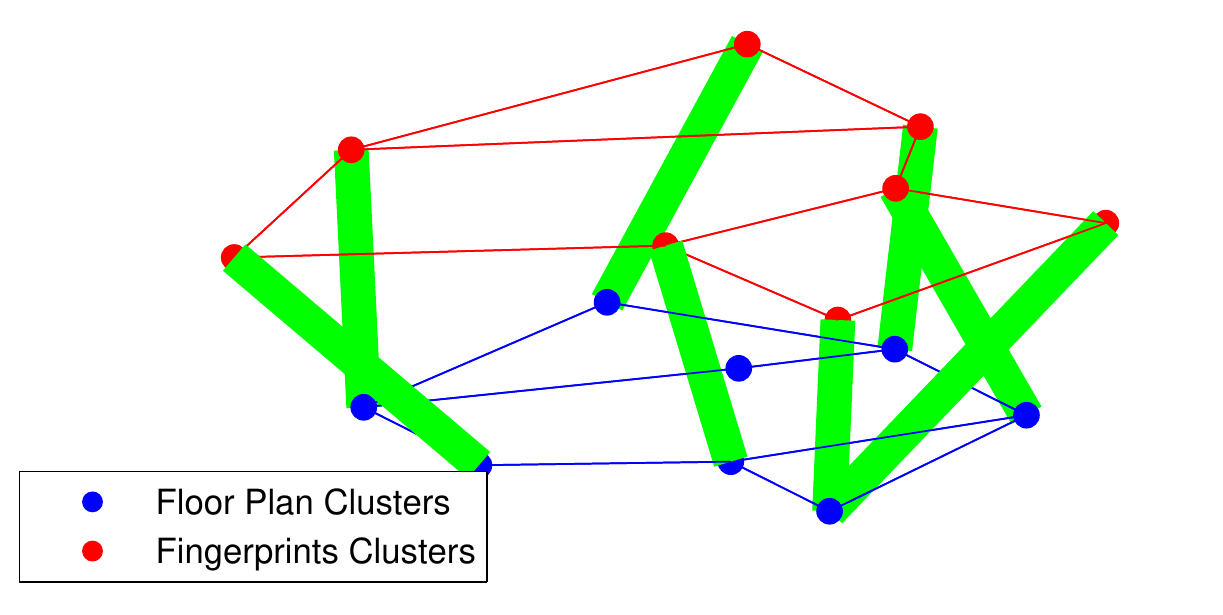}
\vspace{-0.05in}\\
(a) MDS View of Fingerprints, $G^{F}$& (d) Corridor Points of Floor Plan,
$G^{CM}$& (f) Skeletons Matching  \\
\hspace{-0.35in}
\includegraphics[scale=0.45]{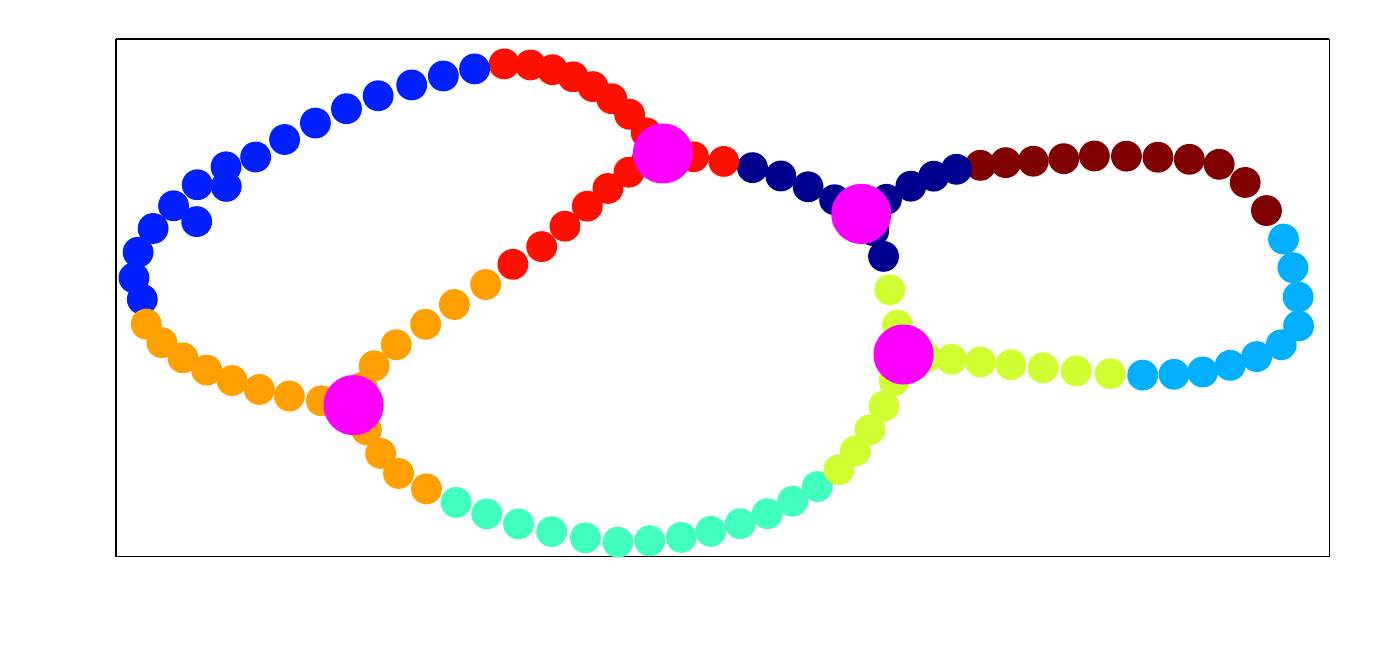}  &
\hspace{-0.35in}
\includegraphics[scale=0.45]{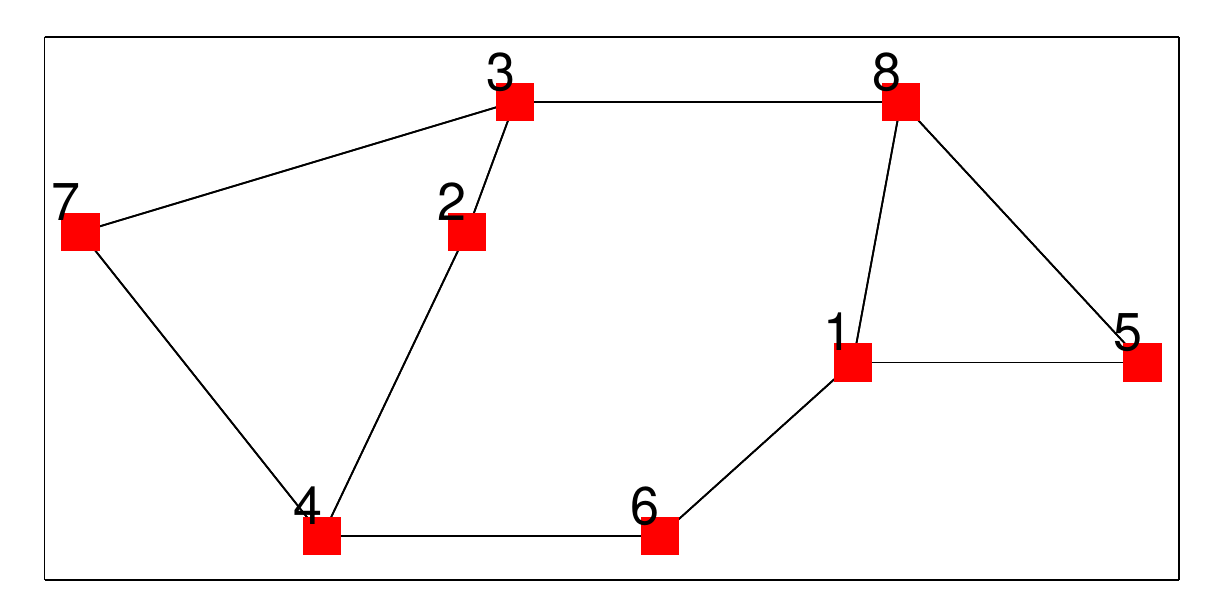} &
\hspace{-0.35in}
 \includegraphics[scale=0.45]{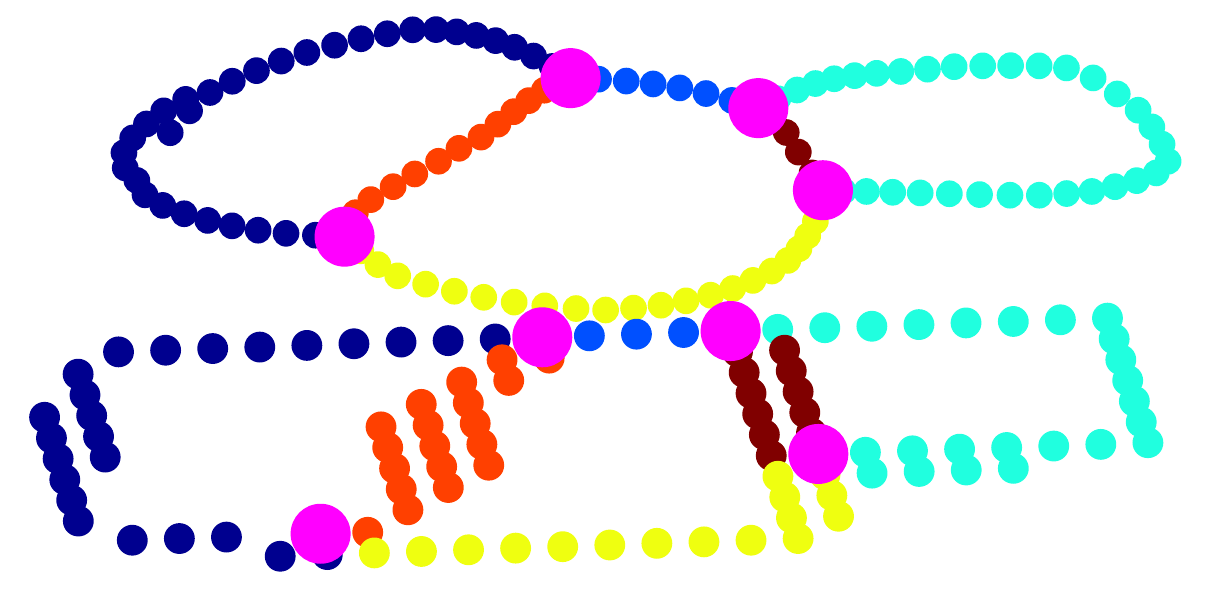} \vspace{-0.05in} \\
(b) Corridor Points of Fingerprints, $G^{CF}$ & (e) Extracted Skeleton Graph,
$G^{SM}$ & (g) Corridor Points Matching \\
 \hspace{-0.35in}
 \includegraphics[scale=0.45]{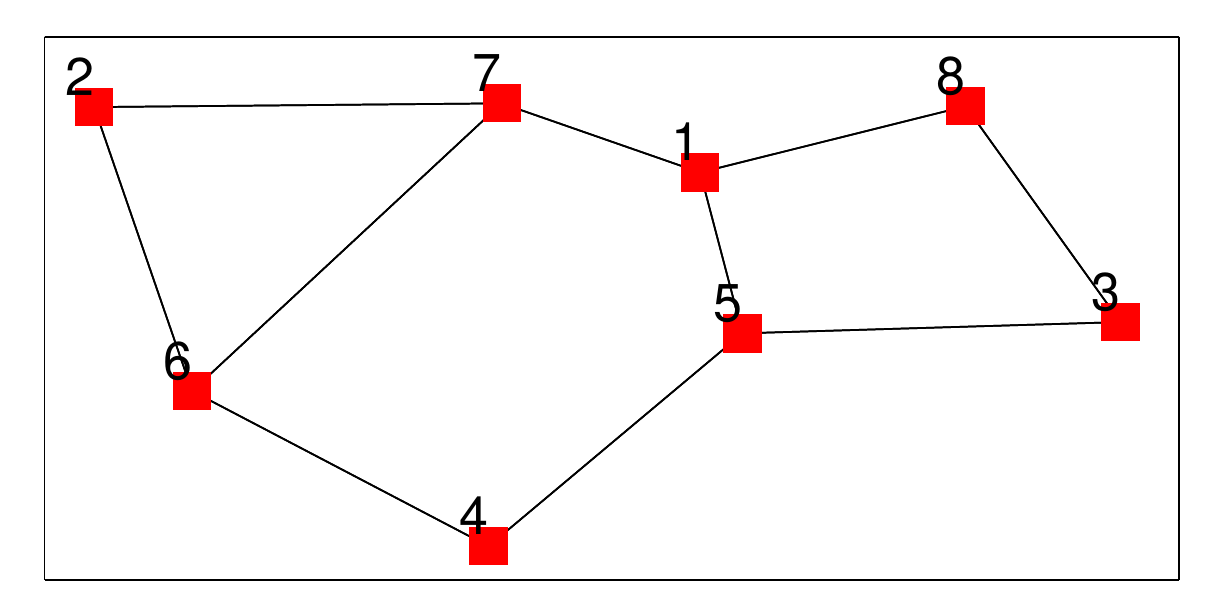} &
 \hspace{-0.35in}
 &
 \hspace{-0.35in}
\includegraphics[scale=0.45]{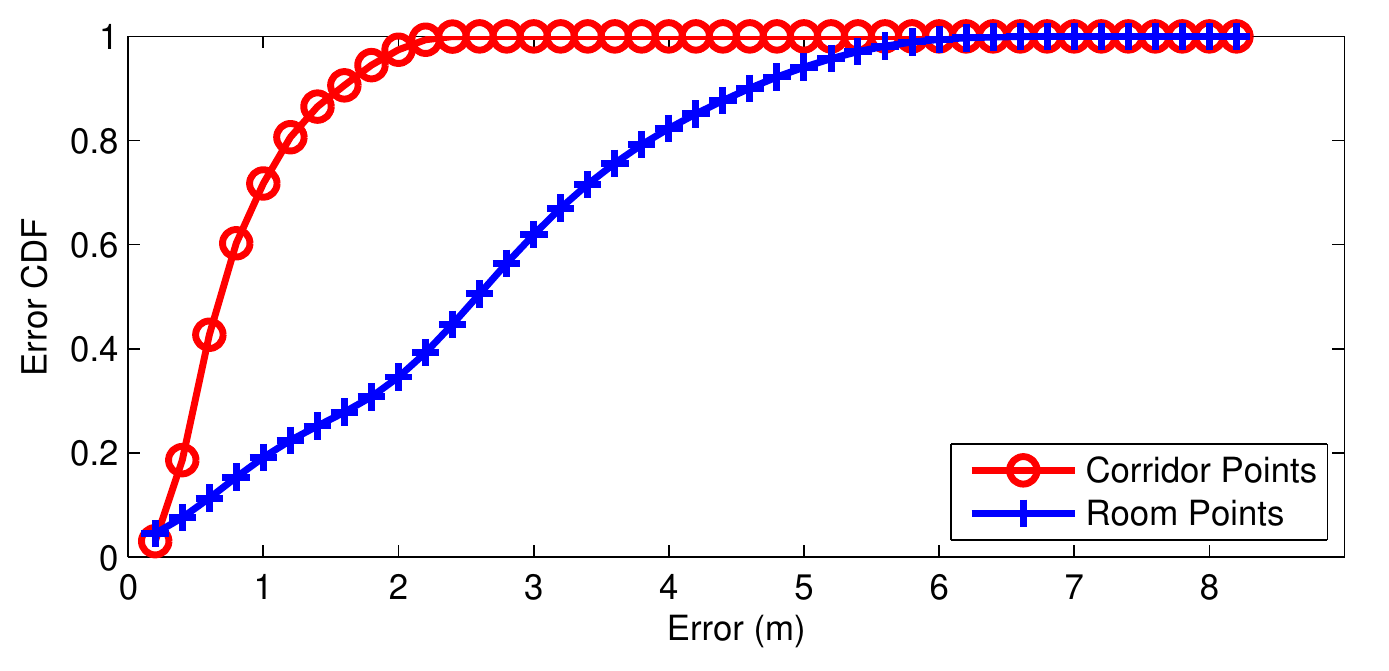} \vspace{-0.05in}\\
\hspace{-0.35in}(c) Extracted Skeleton Graph, $G^{SF}$ &  & (h) Matching Error
CDF\\
\end{tabular}
\end{center}

\caption{Figure (a), (b), and (c) shows the MDS view of fingerprints graph
$G^F$, extracted corridor points graph $G^{CF}$, and skeleton graph $G^{SF}$
respectively. Figure (d), and (e) shows the MDS view of extracted corridor
points graph of floor plan $G^{CM}$, and skeleton graph $G^{SM}$.  Figure (f)
and (g) presents the matching results of skeletons and corridor points. Figure
(h) shows the matching error CDF of both corridor points and room points.  }
 \vspace{-0.15in}
  \label{fig:mapping}
\end{figure*}

During the mapping system test, to fasten the mapping system convergence, 4
students are asked to walk through the environment arbitrarily for 10 minutes,
so as to ensure all rooms and path are covered. During the walking, the android
smart phones in their pockets are constantly communicating with AP to create
enough wireless traffic.

To have a view of extracted fingerprints, we use Multi-Dimensional Scaling (MDS)
algorithm to visualize the fingerprints transition graph $G^F$ in
Fig.~\ref{fig:mapping} (a) and other graph structure, \eg Fig.~\ref{fig:mapping}
(a), (b), (c), and (g).  Fig.~\ref{fig:mapping} (b) presents the extracted
corridor points graph $G^{CF}$  from fingerprints set. To remove all point
except corridor points, the optimal threshold $\tau^*$ is $127$ after 6 round of
attempts. $G^{CF}$ is clustered to extract skeleton graphs, and it is shown in
Fig.~\ref{fig:mapping} (c).

On the other hand, the floor plan is sampled by totally 381 point using
$2\times 2$ grid, and the extract corridor points are shown in
Fig.~\ref{fig:mapping} (d). Despite quite similar to Fig.~\ref{fig:mapping} (c),
they have very different clustering, and the extracted skeleton is presented in
Fig.~\ref{fig:mapping} (e). For both graphs $G^{CF}$ and $G^{CM}$ we detect the
``bridge point'' according to the clustering, the these bridge points are shown
as magenta points in Fig.~\ref{fig:mapping} (b) and Fig.~\ref{fig:mapping} (d).

After graph normalization, Fig.~\ref{fig:mapping} (f) presents the skeleton
matching results. We can see both skeletons are matched even if there are mild
mis-match between Fig.~\ref{fig:mapping} (c) and Fig.~\ref{fig:mapping} (e). The
corridor points are completely matching in Fig.~\ref{fig:mapping} (g) according
to the skeleton matching and the bridge points. The above points are corridor
points from fingerprints, while the bottom points are from floor plan.
Fig.~\ref{fig:mapping} (h) presents the final mapping error. We see more than
80\% of corridor points are matched within an error of 1m, and the largest error
is merely around 2m. The room mapping error is larger. The error for more than
80\% of room points is within 3m, however it is acceptable, since there is no
reference points to align the room points with the room layout.

rooms for 20 minutes to cover all accessible area. The RSS and CSI data is
collected, and Fig.~\ref{fig:mapping} (b) and (b) shows the visualization of
$G^T$ and $G^{CT}$ respectively. Please note that our mapping algorithm requires
only the adjacency matrix of graphs, and the MDS-based position is only for
visualization. The graph matching result is shown in figure (d). For each graph
$G^{SF}$ or $G^{SM}$ there are 8 points, the graph matching under such small
scale could achieve stably and accurate mapping. Figure (e) shows the matching
of corridor points under the guide of skeleton matching and the ''bridge
points''. Figure (f) shows the error of both corridor points mapping and room
points mapping. We can see clear that the corridor points are highly matched.
The corridor points maximum error is still under 2m, while the maximum room
matching error is under 6m.

\subsection{Localization and Tracking}

\begin{figure}[http]
\begin{center}
\includegraphics[scale=0.7]{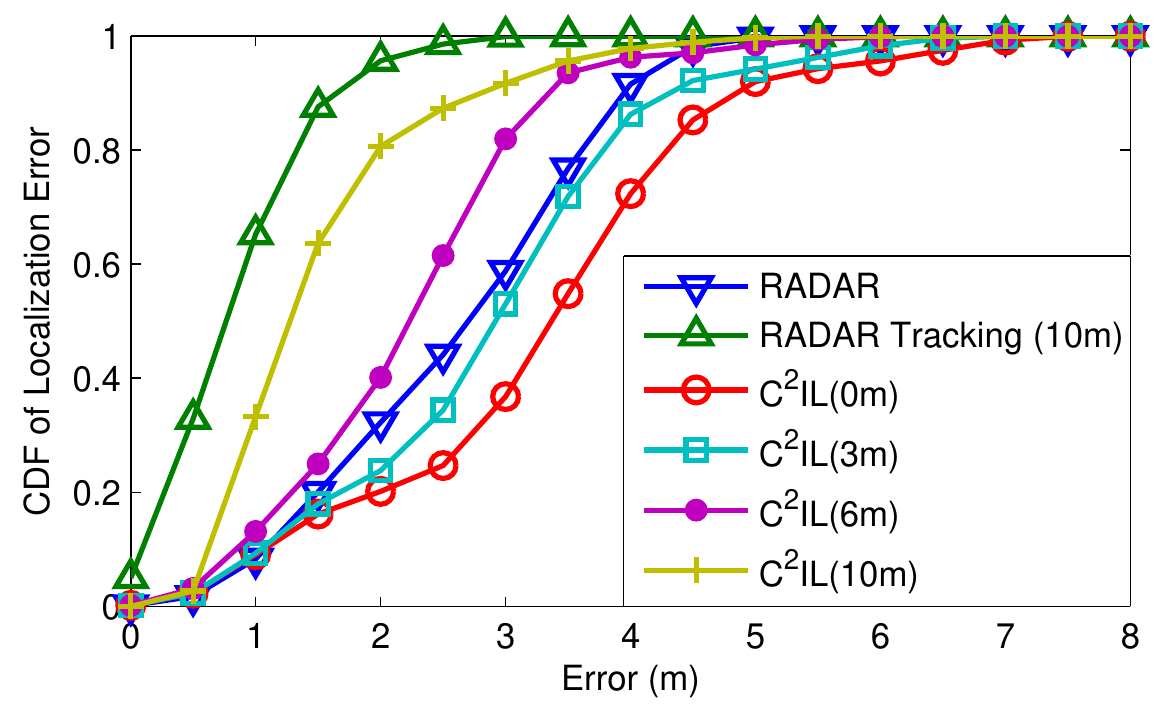}
\end{center}
\vspace{-0.15in}
\caption{The localization \& tracking error CDF with different amount
  $\ell$ of historical RSS fingerprints data. The RADAR scheme is as comparison.
The denotes the direct single point localization, while   denotes the tracking
service with a historical data of 10m.}
\label{fig:tracingcdf}
\vspace{-0.15in}
\end{figure}

In this test, we site-survey the testing environment with approximately $3\times
3$m grid and implement RADAR system as comparison to \ourprotocol. A student
walks along a pre-defined testing trace for 5 rounds, and we randomly select
$\ell$ successive RSS samples for testing.

Fig.~\ref{fig:tracingcdf} shows the error CDF of RADAR and \ourprotocol with
different length of historical data.
Based on the commonly accepted $3\times 3$m grid site-survey,  RADAR has less
than 3.5m error in 80$\%$ situations, while \ourprotocol is only 1m error larger
than RADAR.
When in tracking scenarios, every 3m of increased historical data brings
approximately 0.7m tracking accuracy improvement. When historical data is 10m,
\ourprotocol tracking accuracy is very close to RADAR-based tracking. In
practical tracking scenarios, 10m of historical data is a very low requirement.

\section{Conclusion}
\label{sec:conclusion}

In this paper, we proposed an indoor localization and tracking scheme,
\ourprotocol.
Our scheme does not require using additional sensors, except the
 availability of 802.11n wireless connection.
We believe that \ourprotocol is the first scheme that the really
 benefits from the multipath effect in complex environment. 
An innovative method is proposed to accurately estimate the moving
 speed and distance purely based on 802.11n CSI, which should find a
 wide range of applications alone.
Based on this accurate distance estimation, we built the mapping
 between RSS fingerprints and location using unsupervised learning,
 and  unified the localization and tracking. 
Our extensive evaluation results indicate that our scheme
 \ourprotocol  successfully handles very complex
 indoor structure and simultaneously provides the best performance in
 contribution rate,  localization cost, and localization/tracking
 accuracy.

{{\small
\bibliographystyle{acm}
\bibliography{speedEstimation,location-reference,related-location}

\begin{thebibliography}{10}

\bibitem{ricianKestimate}
{\sc Abdi, A., Tepedelenlioglu, C., Kaveh, M., and Giannakis, G.}
\newblock On the estimation of the k parameter for the rice fading
  distribution.
\newblock {\em Communications Letters, IEEE\/} (2001).

\bibitem{azemi2004mobile}
{\sc Azemi, G., Senadji, B., and Boashash, B.}
\newblock Mobile unit velocity estimation based on the instantaneous frequency
  of the received signal.
\newblock {\em Vehicular Technology, IEEE Transactions on\/} (2004).

\bibitem{azizyan2009surroundsense}
{\sc Azizyan, M., Constandache, I., and Roy~Choudhury, R.}
\newblock Surroundsense: mobile phone localization via ambience fingerprinting.
\newblock In {\em ACM Mobicom'2009}.

\bibitem{bahl2000radar}
{\sc Bahl, P., and Padmanabhan, V.}
\newblock Radar: An in-building rf-based user location and tracking system.
\newblock In {\em IEEE INFOCOM'2000}.

\bibitem{braun1991physical}
{\sc Braun, W., and Dersch, U.}
\newblock A physical mobile radio channel model.
\newblock {\em Vehicular Technology, IEEE Transactions on\/} (1991).

\bibitem{chen1997sbr}
{\sc Chen, S., and Jeng, S.}
\newblock An sbr/image approach for radio wave propagation in indoor
  environments with metallic furniture.
\newblock {\em Antennas and Propagation, IEEE Transactions on\/} (1997).

\bibitem{ez}
{\sc Chintalapudi, K.~K., Iyer, A.~P., and Padmanabhan, V.~N.}
\newblock Indoor localization without the pain.
\newblock In {\em ACM Mobicom'2010}.

\bibitem{cho2010reweighted}
{\sc Cho, M., Lee, J., and Lee, K.}
\newblock Reweighted random walks for graph matching.
\newblock {\em Computer Vision--ECCV\/} (2010).

\bibitem{Clark}
{\sc Clarke, R.~H.}
\newblock A statistical theory of mobile-radio reception.

\bibitem{constandache2010towards}
{\sc Constandache, I., Choudhury, R., and Rhee, I.}
\newblock Towards mobile phone localization without war-driving.
\newblock In {\em IEEE INFOCOM'2010}.

\bibitem{kronecker}
{\sc Garsia, A.~M., and Remmel, J.}
\newblock {Shuffles of permutations and the Kronecker product}.
\newblock {\em Graphs and Combinatorics\/} (1985).

\bibitem{guha2010autowitness}
{\sc Guha, S., Plarre, K., Lissner, D., Mitra, S., Krishna, B., Dutta, P., and
  Kumar, S.}
\newblock Autowitness: locating and tracking stolen property while tolerating
  gps and radio outages.
\newblock In {\em ACM Sensys'2006}.

\bibitem{csitool}
{\sc Halperin, D., Hu, W., Sheth, A., and Wetherall, D.}
\newblock {Predictable 802.11 Packet Delivery from Wireless Channel
  Measurements}.
\newblock In {\em ACM SIGCOMM'2010}.

\bibitem{hill2009electromagnetic}
{\sc Hill, D.}
\newblock {\em Electromagnetic fields in cavities: deterministic and
  statistical theories}.
\newblock Wiley-IEEE Press, 2009.

\bibitem{minkowski}
{\sc Ichino, M., and Yaguchi, H.}
\newblock {Generalized Minkowski metrics for mixed feature-type data analysis}.
\newblock {\em IEEE Transactions on Systems, Man, and Cybernetics\/} (1994).

\bibitem{mds}
{\sc Kruskal, J.~B.}
\newblock {Nonmetric multidimensional scaling: A numerical method}.

\bibitem{leordeanu2005spectral}
{\sc Leordeanu, M., and Hebert, M.}
\newblock A spectral technique for correspondence problems using pairwise
  constraints.
\newblock In {\em IEEE ICCV'2005}.

\bibitem{pushingLimit}
{\sc Liu, H., Gan, Y., Yang, J., Sidhom, S., Wang, Y., Chen, Y., and Ye, F.}
\newblock Push the limit of wifi based localization for smartphones.
\newblock In {\em ACM Mobicom'2012}.

\bibitem{liu2012energy}
{\sc Liu, J., Priyantha, B., Hart, T., Ramos, H., Loureiro, A., and Wang, Q.}
\newblock Energy efficient gps sensing with cloud offloading.

\bibitem{mohanty2005vepsd}
{\sc Mohanty, S.}
\newblock Vepsd: a novel velocity estimation algorithm for next-generation
  wireless systems.
\newblock {\em Wireless Communications, IEEE Transactions on\/} (2005).

\bibitem{centaur}
{\sc Nandakumar, R., Chintalapudi, K.~K., and Padmanabhan, V.~N.}
\newblock Centaur: locating devices in an office environment.
\newblock In {\em ACM Mobicom'2012}.

\bibitem{speedestimation1}
{\sc Pricope, B., and Haas, H.}
\newblock {Experimental Validation of a New Pedestrian Speed Estimator for OFDM
  Systems in Indoor Environments}.
\newblock In {\em IEEE GLOBECOM'2011}.

\bibitem{Zee}
{\sc Rai, A., Chintalapudi, K.~K., Padmanabhan, V.~N., and Sen, R.}
\newblock Zee: zero-effort crowdsourcing for indoor localization.
\newblock In {\em ACM Mobicom'2012}.

\bibitem{rappaport1996wireless}
{\sc Rappaport, T.}
\newblock {\em Wireless communications: principles and practice}.
\newblock IEEE press, 1996.

\bibitem{spinloc}
{\sc Sen, S., Choudhury, R.~R., and Nelakuditi, S.}
\newblock Spinloc: spin once to know your location.
\newblock In {\em ACM HotMobile'2012}.

\bibitem{sen2012you}
{\sc Sen, S., Radunovic, B., Choudhury, R., and Minka, T.}
\newblock You are facing the mona lisa: spot localization using phy layer
  information.
\newblock In {\em ACM Mobisys'2012}.

\bibitem{pinloc}
{\sc Sen, S., Radunovic, B., Choudhury, R.~R., and Minka, T.}
\newblock You are facing the mona lisa: spot localization using phy layer
  information.
\newblock In {\em ACM Mobisys'2012}.

\bibitem{sklar1997rayleigh}
{\sc Sklar, B.}
\newblock Rayleigh fading channels in mobile digital communication systems. i.
  characterization.
\newblock {\em Communications Magazine, IEEE\/} (1997).

\bibitem{von2007tutorial}
{\sc Von~Luxburg, U.}
\newblock A tutorial on spectral clustering.
\newblock {\em Statistics and computing, Springer\/} (2007).

\bibitem{FILA}
{\sc Wu, K., Xiao, J., Yi, Y., Gao, M., and Ni, L.~M.}
\newblock {FILA: Fine-grained indoor localization}.
\newblock In {\em IEEE INFOCOM'2012}.

\bibitem{lifs}
{\sc Yang, Z., Wu, C., and Liu, Y.}
\newblock Locating in fingerprint space: wireless indoor localization with
  little human intervention.
\newblock In {\em ACM Mobicom'2012}.

\bibitem{youssef2008horus}
{\sc Youssef, M., and Agrawala, A.}
\newblock The horus location determination system.
\newblock {\em Wireless Networks\/} (2008).

\bibitem{zhang2011antenna}
{\sc Zhang, Z., Zhou, X., Zhang, W., Zhang, Y., Wang, G., Zhao, B., and Zheng,
  H.}
\newblock I am the antenna: Accurate outdoor ap location using smartphones.
\newblock In {\em ACM Mobicom'2011}.

\bibitem{zonoozi1996shadow}
{\sc Zonoozi, M., and Dassanayake, P.}
\newblock Shadow fading in mobile radio channel.
\newblock In {\em IEEE PIMRC'1996}.

\end{thebibliography}
}}

\end{document}